\documentclass[journal]{IEEEtran}
\ifCLASSINFOpdf
\else
\fi
\usepackage{bm}
\usepackage{amsmath}
\usepackage{booktabs}
\usepackage{indentfirst}
\usepackage{amsfonts}
\usepackage{graphicx}
\usepackage{subfigure}
\usepackage{cite}
\usepackage{float}
\usepackage{hyperref}
\usepackage{makecell}
\usepackage{epstopdf}
\usepackage{soul}
\usepackage{footnote}
\usepackage{xcolor}
\hypersetup{hypertex=true,
colorlinks=true,
linkcolor=black,
anchorcolor=black,
citecolor=black}

\begin{document}
%
\title{A Holistic Robust Motion Control Framework for Autonomous Platooning}
%
%
%

\author{Hong Wang,
        Li-Ming Peng,
        Zichun Wei,
        Kai Yang,
        Xian-Xu ‘Frank’ Bai,
        Luo Jiang,
        and~Ehsan Hashemi*
\thanks{This work was supported by National Key R\&D Program of China:2020YFB1600303, and the National Science Foundation of China Project: U1964203 and 52072215.(The first two authors Hong Wang and Li-Ming Peng contributed equally to this work. Corresponding author:  Ehsan Hashemi).}
\thanks{H. Wang is with the School of Vehicle and Mobility, Tsinghua University, Beijing, 100084, China.(email: hong\_wang@tsinghua.edu.cn).}
\thanks{Li-Ming Peng and Xian-Xu ‘Frank’ Bai are with the department of Vehicle Engineering, Hefei University of Technology, Hefei, 230009, China (e-mail: alaric.peng@gmail.com, ~e-mail: bai@hfut.edu.cn)}
\thanks{Zichun Wei is with the Viterbi school of engineering, University of Southern California, Los Angeles, 90089, United States of America. (e-mail: zichunwe@usc.edu)}
\thanks{K. Yang is with College of Mechanical and Vehicle Engineering, Chongqing University, Chongqing 400044, China (email: kaiyang0401@gmail.com).}
\thanks{Luo Jiang and Ehsan Hashemi are with Department of Mechanical Engineering, University of Alberta, Edmonton, AB, Canada T6G 1H9. (e-mail: luo.jiang@ualberta.ca,~e-mail: ehashemi@ualberta.ca)}}


%
%

\markboth{IEEE Internet of Things Journal}%
{Shell \MakeLowercase{\textit{et al.}}: Bare Demo of IEEEtran.cls for IEEE Journals}
%



\maketitle

\begin{abstract}
Safety is the foremost concern for autonomous platooning. The vehicle-to-vehicle (V2V) communication delay and the sudden appearance of obstacles will trigger the safety of the intended functionality (SOTIF) issues for autonomous platooning. This research proposes a holistic robust motion controller framework (MCF) for an intelligent and connected vehicle platoon system. The MCF utilizes a hierarchical structure to resolve the longitudinal string stability and the lateral control problem under the complex driving environment and time-varying communication delay. Firstly, the H-infinity feedback controller is developed to ensure the robustness of the platoon under time-varying communication delay in the upper-level coordination layer (UCL). The output from UCL will be delivered to the lower-level motion-planning layer (LML) as reference signals. Secondly, the model predictive control (MPC) algorithm is implemented in the LML to achieve multi-objective control, which comprehensively considers the reference signals, the artificial potential field, and multiple vehicle dynamics constraints. Furthermore, three critical scenarios are co-simulated for case studies, including platooning under time-varying communication delay, merging, and obstacle avoidance scenarios. The simulation results indicate that, compared with single-structure MPC, the proposed MCF can offer a better suppression on position error propagation, and get improvements on maximum position error in the three scenarios by $19.2\%$, $59.8\%$, and $15.3\%$, respectively. Last, the practicability and effectiveness of the proposed MCF are verified via hardware-in-the-loop experiment. The average conducting time of the proposed method on Speedgoat real-time target machine is 1.1 milliseconds, which meets the real-time requirements.
\end{abstract}


%
\IEEEpeerreviewmaketitle

\section{Introduction}
%
%
%
%
\subsection{Motivation}
\IEEEPARstart{T}{he} intelligent and connected vehicle (ICV) technique plays a critical role in the future intelligent transportation system (ITS). It provides effective solutions to deal with the existing traffic problems, such as traffic accidents and traffic jams\cite{lee2012development, zhao2016integrated}. One of the typical applications of the ICV technique is the autonomous platooning. Compared with the automation of a single vehicle, the platooning control of ICVs is currently applied more extensively on the transportation missions. These missions have the characteristics of high degrees of specialization, independent scenarios, and a narrow operational design domain (ODD), such as mining trucks, port transporters, and heavy-duty vehicles on the freeway \cite{shladover1991automated, gao2016empirical}. Platooning indicates that the vehicles should follow the velocity of the leading vehicle while keeping a limited inter-vehicle space to improve the traffic efficiency and energy consumption  \cite{hong2020joint}. However, the limited inter-vehicle spaces add pressure on the guarantee of safety for autonomous platooning \cite{ju2020deception}. And the safety of autonomous platoon suffers great risks from the insufficiency of intended functionality in complex operating environment, since ICVs rely on the complex perception, control, and communication systems with uncertainty to sense the environment with complex disturbance and generate control commands \cite{wang2018review}.
\begin{figure}
    \centering
    \includegraphics[width=9cm]{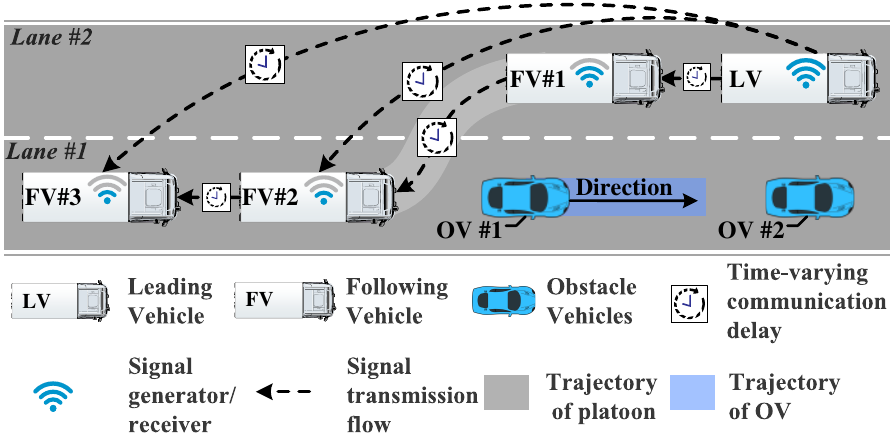}
    \vspace{-0.6cm}
    \caption{Critical scenarios with time-varying communication delay and obstacle avoidance.}
    \vspace{-0.5cm}
    \label{Fig1}
\end{figure}
\par Compared with the functional safety issues related to the malfunction of electrical/electronic (E/E) systems, the safety of the intended functionality (SOTIF) refers to the absence of unreasonable risks due to hazards that result from insufficiencies of the intended functionality or reasonably foreseeable misuse \cite{iso2019pas}. According to the system theoretic process analysis (STPA) accomplished by the EU project ENSEMBLE \cite{konstantinopoulou2019specifications}, there are several types of SOTIF challenges for autonomous platooning such as the infrastructure problems, the misuses of human drivers, communication problems and external disturbance, etc. In particular, the communication and external disturbance challenges will jeopardize the SOTIF in longitudinal and lateral control, which is essential for autonomous platooning \cite{wang2022robust}. For instance, Fig. \ref{Fig1} depicts a critical scenario for autonomous platooning, where the platoon needs to avoid obstacle vehicles with the vehicle-to-vehicle (V2V) communication. The V2V communication delay and the sudden appearance of obstacles may cause collisions in autonomous platoon due to the functional insufficiencies of the longitudinal and lateral control. The main focus of this research is to design a controller framework that guarantees the safety of the ICV platoon in corresponding critical scenarios by collaborative longitudinal and lateral control.
\subsection{Related Research}
The longitudinal string stability of a platoon means that the perturbation of the leading vehicle should not be amplified while propagating along the string. Significant research efforts have been continuously devoted to maintaining longitudinal string stability for autonomous platooning. It was pointed out in \cite{seiler2004disturbance} that the conventional adaptive cruise control with the predecessor following (PF) topology could not guarantee the string stability. The development of vehicle-to-everything (V2X) technique provides the possibility of sharing the state information among the vehicles \cite{li2019platoon, zhao2020vehicle}. For instance, a data-driven model predictive control (MPC) method is proposed with the help of V2V wireless communication, which successfully maintains stability of mixed vehicle platoon \cite{lan2021data}. However, the communication delay and packet dropouts occurring during the V2X process will jeopardize the string stability of the platoon and even cause vehicle collisions \cite{zhao2020stability}.
\par The influence of the communication delay to the robustness of vehicle platoons was analyzed and considered in many research \cite{mamduhi2020event}. An H-infinity controller based on the Lyapunov-Krasovskii approach was proposed for the string stability of the heterogeneous vehicle platoon in \cite{xu2018stable} to deal with the communication delay. Zheng \textit{et al}. \cite{zheng2016distributed} proposed a distributed MPC based on unidirectional communication topology to ensure the string stability of heterogeneous platoon. The authors adopted the sum of each vehicle’s cost function as the Lyapunov candidate to theoretically verify its stability. A constant and homogeneous communication delay was considered in the existing studies. However, the communication delay tends to be heterogeneous and time-varying in practice due to the vehicle speed and communication density \cite{xu2017dsrc}. Therefore, it is more appealing to develop a controller for autonomous platoon that considers the heterogeneous and time-varying communication delay.
\par Besides, while the conventional single-structure robust controllers can maintain the longitudinal string stability in specific scenarios, they are still lack of the considerations of the obstacle avoidance and vehicle dynamics limitations. In fact, autonomous platooning is a multi-objective problem. When the driving safety of the platoon confronts complicated scenarios, it is inadequate to only maintain the longitudinal string stability\cite{chen2020cooperative}. The lateral control should be implemented while simultaneously considering the obstacles and vehicle dynamic limitations \cite{baldi2020establishing}, which is not yet well addressed in the available research literature. For the automation of a single vehicle, the driving safety is guaranteed by the optimal motion planning. The common algorithms used for motion planning include the graph-based methods such as the A-star algorithm \cite{montemerlo2008junior}, the meta-heuristic optimization methods such as genetic algorithm and simulated annealing \cite{zheng2005evolutionary}, and the artificial potential field (APF) method \cite{ji2016path}. Optimization-based approaches are among the state-of-the-art approaches in the field of motion planning for autonomous vehicles \cite{liu2015predictive}. Among these approaches, the MPC has been well studied and applied due to its ability to handle multiple constraints \cite{yang2021comparative,huang2018review}. For example, the APF-based MPC controller was used to calculate the optimal trajectory with the multi-objective constraints in \cite{rasekhipour2016potential}, comprehensively considering the obstacle in the ODD.
\par However, an effective solution to the motion planning problem for autonomous platooning still remains a challenge in the case of platooning control. When the vehicles move simultaneously as a platoon, the motion planning will encounter issues, e.g. high occupation of lane space and low mobility. As the over-taking and merging scenarios occur frequently, the lane space requirement and the risk will both be high if the vehicles change the lane simultaneously. Therefore, the distributed controllers should be deployed, which also means that the lateral control of autonomous vehicles in the platoon should be independent while the longitudinal control is coordinated to guarantee the longitudinal string stability. The above discussion demonstrates that the conventional single-structure control strategies are insufficient for the comprehensive requirements of platooning control. As a result, this paper presents a motion controller framework to resolve aforementioned issues for autonomous platoon.
\subsection{Contribution}
In brief, the main contributions of this paper are as follows:
\par (i) A H-infinity feedback controller is designed to guarantees the longitudinal string stability of autonomous platoon, subject to time-varying communication delay, time delay in propulsion, and dynamic fluctuation of leading vehicle. The time-varying communication delay, modeled as a system uncertainty, is more realistic than the constant and homogeneous delay used in the existing literature. Moreover, the first-order approximation of acceleration response captures more realistic vehicle dynamics than the point-mass model used in the existing literature.
\par (ii) The APF-based distributed MPC controllers are designed to achieve asynchronous motion planning in critical scenarios, which fill in gaps of the lateral control of autonomous platooning in the existing literature.
\par (iii) A holistic robust motion controller framework is proposed to achieve collaborative longitudinal and lateral control for autonomous platooning, which has not been explored in existing literature yet. The string stability and lateral safety are balanced by comprehensively considering the reference signal from H-infinity controller, the artificial potential field, and multiple vehicle dynamics constraints. Moreover, the Hardware-in-the-loop experiment is conducted to verify the effectiveness and real-time capacity of the proposed framework.
\subsection{Paper Organization}
The remaining of this paper is organized as follows. In Section II, the H-infinity feedback controller is designed and analyzed. The APF-based distributed MPC controllers are also presented in this section. Furthermore, the holistic robust motion controller framework (MCF) is modeled and discussed. In Section III, the case studies are carried out, and the performance of the proposed MCF is analyzed with corresponding scenarios. Moreover, the comprehensive capacity of the proposed MCF and the promising improvements are discussed. The experimental verification of the proposed framework is given in Section IV. The conclusion and future work are demonstrated in Section V.
\section{Holistic robust motion controller framework}
The design of holistic robust motion controller framework for autonomous platooning is introduced in this section. The process includes the design of H-infinity feedback controller, the MPC algorithm for motion planning, and the introduction of the framework for holistic control requirements.
\subsection{H-infinity Feedback Controller Design}
 The longitudinal dynamics of a vehicle is inherently nonlinear in reality, including the complex characteristics of the engine/motor, the air drag force and the tire adhesion conditions, etc. \cite{gao2016robust}. The proposed platoon state space model of longitudinal dynamics uses the first-order approximation to estimate the acceleration response \cite{li2017distributed} as follows, by which the effect of time delays in propulsion is considered.
 \begin{equation}
\label{Eq1}
\begin{aligned}
    {\dot{a}}_{i} &= \frac{1}{\tau_i}(u_{i} - a_{i})
\end{aligned}
\end{equation}
\par Moreover, based on the constant spacing policy, the vehicle dynamic model is transformed to the error model compared with the leading vehicle as follows:
\begin{equation}
\label{Eq2}
\begin{aligned}
    {\hat{s}}_{i} &= s_{0} - s_{i} - iD_{i}\\
    {\hat{v}}_{i} &= v_{0} - v_{i} \\
    {\hat{a}}_{i} &= a_{0} - a_{i}
\end{aligned}
\end{equation}
    where $D_i$ is the desired distance, $s_0$, $v_0$ and $a_0$ denote the actual position, velocity and acceleration of the leading vehicle, respectively, and $s_i$, $v_i$ and $a_i$ represent the actual position, velocity and acceleration of the $i^{th}$ following vehicle in the platoon, respectively. The position, velocity and acceleration errors between the leading vehicle and the $i^{th}$ following vehicle are denoted by ${\hat{s}}_{i}$, and ${\hat{v}}_{i}$ and ${\hat{a}}_{i}$, respectively.
\par Based on the system state vectors $\bm{{x}}_{Hi}=[{\hat{s}}_{i}, {\hat{v}}_{i}, {\hat{a}}_{i}]^T$, the vehicle state space model can be defined as follows:
\begin{equation}
\label{Eq3}
    \bm{\dot{{x}}}_{Hi} = \bm{A}_i\bm{{x}}_{Hi}+\bm{B}_{ui}\bm{u}_i+\bm{B}_{wi}\bm{w}_i
\end{equation}
In (3), $\bm{A}_i$, $\bm{B}_{ui}$, $\bm{B}_{wi}$ denote the system matrix, control matrix and disturbance matrix of the $i^{th}$ vehicle in the platoon, respectively, which are given as:
\par
    $\bm{A}_i= \left [
    \begin{array}{ccc}
         0 & 1 & 0  \\
         0 & 0 & 1  \\
         0 & 0 & -1/\tau_i 
    \end{array} \right ]$,
    $\bm{B}_{ui}= \left [
    \begin{array}{ccc}
         0 & 0 & -1/\tau_i 
    \end{array} \right ]^T$, 
\par
$\bm{B}_{wi}= \left [
    \begin{array}{ccc}
         0 & 0 & 1/\tau_i  \\
         0 & 0 & 1
    \end{array} \right ]^T$
\\
\par \noindent and $\bm{u}_i$ and $\bm{w}_i$ are the control input and disturbance, respectively. The acceleration ${{a}}_{0}$ of the leading vehicle is regarded as a disturbance, since that it is an external input drifting the platoon system away from the equilibrium position. $\tau_i$ is the time delay in propulsion. The state space model of each vehicle in the platoon is consistent because the platoon is considered a homogeneous system.
\par To realize the string stability of the platoon, the linear control law is chosen for every following vehicle under the predecessor-leader following (PLF) topology. Considering the time-varying communication delay as a system uncertainty occurred in the PLF topology, the control gain $\bm{u}_i$ is designed as follows:
\begin{equation}
\label{Eq4}
\begin{aligned}
    &\bm{u}_i=\Bar{K}_{1i}\bm{x}_{Hi}(t-\lambda_t)+\Bar{K}_{2i}\bm{x}_{H(i-1)}(t-\lambda_t) \\ 
    &s.t.~~\lambda_t \in [0, h_1], \dot{\lambda}_t < \mu_1<1
\end{aligned}
\end{equation}
where $\Bar{K}_{1i}$ and $\Bar{K}_{2i}$ denote the control gain matrix of the $i_{th}$ vehicle, $\bm{x}_{Hi}(t-\lambda_t)$ and $\bm{x}_{H(i-1)}(t-\lambda_t)$ refer to the system state vectors under the communication delay, and $\lambda_t$, $h_1$ and $\mu_1$ represent the communication delay, upper bound of the communication delay and the upper bound of the derivative of the communication delay, respectively.
\par Considering the platoon state vector $\bm{X}=[\bm{x}_{H1}, \bm{x}_{H2}, ..., \bm{x}_{Hn}]^T$, the platoon control input can be defined as follows:
\begin{equation}
\label{Eq5}
\begin{aligned}
    \bm{U}=\bm{K}\bm{X}(t-\lambda_t)
\end{aligned}
\end{equation}
\par Moreover, the platoon state space model is constructed as follows:
\begin{equation}
\label{Eq6}
\begin{aligned}
    \dot{\bm{X}}(t)&=\bm{A}\bm{X}(t)+\bm{B}_u\bm{K}\bm{X}(t-\lambda_t)+\bm{B}_w\bm{W}(t)\\
    \bm{Z}(t)&=\bm{C}\bm{X}(t)\\
    \bm{X}(t)&=\bm{\emptyset}(t)~~~(t \in [-\infty, 0])
\end{aligned}
\end{equation}
where $\bm{Z}(t)=[\hat{s}_1, \hat{s}_2, ..., \hat{s}_n]^T$, $\bm{C}=\bm{I}_n\bigotimes[1,0,0]$, $\bm{A}$, $\bm{B}_u$, $\bm{B}_w$ denote the platoon system matrix, control matrix and disturbance matrix, respectively. All these matrices are diagonal matrices, and omitted here for simplicity. The initial status of the platoon is $\bm{\emptyset}(t)$, $\bm{Z}(t)$ represents the output matrix, $\bm{I}_n$ refers to the n-dimensional identity matrix, and the symbol $\bigotimes$ denotes the Kronecker product.
\par The robustness of a system to uncertainties and external disturbances is a foundational system-theoretic concept, and the H-infinity norm plays a substantial role in uncertainty modelling and robust controller synthesis \cite{pare2019networked}. For the platoon state space model shown in (6), the performance index $J\left(\bm{W}\right)$ is defined as:
\begin{equation}
\label{Eq7}
    \begin{aligned}
    J(\bm{W}) = {\int_0^\infty}[\bm{Z}^T(t)\bm{Z}(t)-\gamma^2\bm{W}^T(t)\bm{W}(t)]dt
    \end{aligned}
\end{equation}
where $\gamma$ denotes the performance index of the H-infinity controller.
\par \noindent \textbf{\textit{Theorem 1}}: For the given time-varying delay $(\lambda_t \in [0,h_1] ,\quad  \Dot{\lambda}_t<\mu_1<1)$ and scalar $\gamma \ge 0$, the platoon system is asymptotically stable and satisfies $J(\bm{W})<0$, if there exist matrices $\bm{\bar{P}}>0$, $\bm{\bar{Z}}>0$, $\bm{\bar{Q}}>0$ and any appropriately dimensioned matrices $\bm{Y}$ and $\bm{I}$ (identity matrix) such that the following inequality is satisfied:
\begin{equation}
\label{Eq8}
   \begin{aligned}
    \left[\begin{array}{ccccc} 
    \bm{\Lambda_1}&\bm{B_u}\bm{Y}+\frac{\bm{\bar{Z}}}{h_1}&\bm{B_w}&\bm{\bar{P}}\bm{A}^T&\bm{\bar{P}}\bm{C}^T \\
    *&-\bm{\bar{Q}}-\frac{\bm{\bar{Z}}}{h_1}&0&(\bm{B_u}\bm{Y})^T&0\\
    *&*&-\gamma^2\bm{I}&\bm{B_w}^T&0\\
    *&*&*&-\frac{\bm{\bar{P}}\bm{\bar{Z}}^{-1}\bm{\bar{P}}}{h_1}&0 \\
    *&*&*&*&-\bm{I}
    \end{array}\right]<0
    \end{aligned}
\end{equation}
where $\bm{\Lambda}_1 = \bm{A\bar{P}}+\bm{\bar{P}A}^T+\frac{\bm{\bar{Q}}}{1-\mu_1}-\frac{\bm{\bar{Z}}}{h_1}$.
\par \noindent \textbf{\textit{Proof}}: Based on the Lyapunov stability theorem, the Lyapunov-Krasovskii functional is defined as
\begin{equation}
\label{Eq9}
    \begin{aligned}
    V(\bm{X}_t)&=\bm{X}^T(t)\bm{P}\bm{X}(t)+\frac{1}{1-\mu_1}\int_{t-\lambda_t}^t\bm{X}^T(s)\bm{Q}_1\bm{X}(s)ds\\
    &+\int_{t-h_1}^t(s+(h_1-t))\Dot{\bm{X}}^T(s)\bm{Z}_1\Dot{\bm{X}}(s)ds
    \end{aligned}
\end{equation}
where $\bm{P}$, $\bm{Q}_1$ and $\bm{Z}_1$ are positive definite matrices to be found.
\par Furthermore, the H-infinity controller can fulfill the desired performance index, if it satisfies the following condition:
\begin{equation}
\label{10}
    \begin{aligned}
    \frac{{\lVert \bm{Z}(t) \rVert}_\infty}{{\lVert \bm{W}(t) \rVert}_\infty} < \gamma
\end{aligned}
\end{equation}

\par Using linear transformation and Schur Complementary Lemma, the condition given in (10) is transformed as follows:
\begin{equation}
\label{Eq11}
    \begin{aligned}
    \bm{Z}^T(t)\bm{Z}(t)-\gamma^2\bm{W}^T(t)\bm{W}(t)+\Dot{V}(X_t) <\bm{\varepsilon}^T\bm{\Phi}_1\bm{\varepsilon} 
\end{aligned}
\end{equation}
\par $\bm{\varepsilon} = [\bm{X}^T(t), \bm{X}^T(t-\lambda_t), \bm{W}^T(t)]^T$

\begin{equation}\nonumber
    \begin{aligned}
    \bm{\Phi}_1=\left[\begin{array}{ccccc}
    \bm{\Lambda}_2&\bm{P}\bm{B}_u\bm{K}+\frac{\bm{Z}_1}{h_1}&\bm{PB}_w&\bm{A}^T&\bm{C}^T\\
    *&-\bm{Q}_1-\frac{\bm{Z}_1}{\bm{h}_1}&0&(\bm{B}_u\bm{K})^T&0 \\
    *&*&-\gamma^2\bm{I}&\bm{B}_w^T&0 \\
    *&*&*&-\frac{\bm{Z}_1^{-1}}{h_1}&0\\
    *&*&*&*&-\bm{I}
    \end{array}\right]
\end{aligned}
\end{equation}
where $\bm{\Lambda}_2 = \bm{PA}+\bm{A}^T\bm{P}+\frac{1}{1-\mu_1}\bm{Q}_1-\frac{\bm{Z}_1}{h_1}$.
\par The inequality given in (11) can be satisfied when $\bm{\Phi}_1 < 0$. The $diag(\bm{P}^{-1},\bm{P}^{-1},\bm{I},\bm{I},\bm{I})$ is used to multiply on the both sides of $\bm{\Phi}_1$,  and let $\bm{\bar{P}}=\bm{P}^{-1}$, $\bm{Y}=\bm{KP}^{-1}$, $\bm{\bar{Z}}=\bm{P}^{-1}\bm{Z}_1\bm{P}^{-1}$ and $\bm{\bar{Q}}=\bm{P}^{-1}\bm{Q}_1\bm{P}^{-1}$. Subsequently, the necessary and sufficient condition given in (\ref{Eq8}) can be formulated, and thus \textbf{\textit{Theorem 1}} is proved.
\par The matrix inequalities cannot be computed directly due to the nonlinear elements in (\ref{Eq8}). The transformation algorithm in \cite{moon2001delay} is used to transform the original non-linear matrices inequality to nonlinear optimization problem constrained by linear matrix inequalities. The non-linear element in (8) is replaced by $\bm{\bar{S}}=\bm{\bar{P}}\bm{\bar{Z}}^{-1}\bm{\bar{P}}$ to get $\bm{\Phi}_2$. The non-linear optimization problem can be represented as follows:
\begin{equation}
\label{Eq12}
\begin{aligned}
    & min \quad trace\{\bm{\bar{S}}\bm{\bar{T}}+\bm{\bar{P}}\bm{\bar{J}}+\bm{\bar{Z}}\bm{\bar{R}}\}\\
    &s.t. \quad \bm{\Phi}_2 < 0;\quad \bm{\bar{P}}<0;\quad \bm{\bar{Q}}<0;\quad \bm{\bar{Z}}<0\\
    &\left[\begin{array}{cc}
    \bm{\bar{T}}&\bm{\bar{J}}\\
    *&\bm{\bar{R}}
    \end{array}\right]\geq 0 \quad 
    \left[\begin{array}{cc}
    \bm{\bar{S}}&\bm{I}\\
    *&\bm{\bar{T}}
    \end{array}\right]\geq 0\\
    &\left[\begin{array}{cc}
    \bm{\bar{P}}&\bm{I}\\
    *&\bm{\bar{J}}
    \end{array}\right]\geq 0 
    \quad \left[\begin{array}{cc}
    \bm{\bar{Z}}&\bm{I}\\
    *&\bm{\bar{R}}
    \end{array}\right]\geq 0\\
\end{aligned}
\end{equation}
where $\bm{\bar{T}}=\bm{\bar{S}}^{-1}$, $\bm{\bar{J}}=\bm{\bar{P}}^{-1}$, $\bm{\bar{R}}=\bm{\bar{Z}}^{-1}$.
\par The non-linear optimization problem (12) can be solved to obtain the optimized solution of $\bm{\bar{P}}$, $\bm{\bar{Y}}$, $\bm{\bar{Q}}$, $\bm{\bar{Z}}$ via the Cone Complementarity Linearization method in \cite{el1997cone}. Furthermore, the control gain $\bm{K=\bar{Y}\bar{P}}$ can be computed.
\par \noindent \textbf{\textit{Remark 1:}} The Cone Complementarity Linearization and non-linear optimization are implemented via MATLAB and MOSEK toolbox. Since that the scale of (8) increases when the number of following vehicle increases, the Lyapunov-Razumikhin functional is a better method than the Lyapunov-Krasovskii functional in resolving the control problem of large-scale networks, which is also used when the system delay is unknown or rapidly varying \cite{wu2006delay}. To enhance the scalability of the proposed method, we will study the stability of autonomous platoon based on the Razumikhin theorem in the future work. Furthermore, the increasing complexity of Lyapunov functional candidate tends to decrease the conservativeness of the system stability conditions at the expense of increased computational complexity.
\par \noindent \textbf{\textit{Lemma  1:}} Theoretically let $\dot{\bm{\xi}}=\bm{F}(t,\bm{\xi})$:
\par (i) The stable equilibrium point $\bm{F}(t_{0},0)=0$ is asymptotically stable if for each $\bm{\epsilon}>0$, there is  $\eta=\eta(\bm{\epsilon},t_{0})>0$ such that $\|\bm{\xi}(t_{0})\|< \eta \Rightarrow \|\bm{\xi}(t)\|<\bm{\epsilon}$, $\forall t \geq t_{0} \geq 0$; and there is $c=c(t_{0})>0$ such that $\lim\limits_{t \to \infty}\bm{\xi}(t) = 0$, for all $\|\bm{\xi}(t_{0})\|<c$.
\par (ii) The stable equilibrium point $\bm{F}(t_{0},0)=0$ is uniformly asymptotically stable if for each $\bm{\epsilon}>0$, there is $\eta=\eta(\bm{\epsilon})>0$, which is independent of $t_{0}$ such that $\|\bm{\xi}(t_{0})\|< \eta \Rightarrow \|\bm{\xi}(t)\|<\bm{\epsilon}$, $\forall t \geq t_{0} \geq 0$; and there is $c>0$, which is also independent of $t_{0}$, such that $\lim\limits_{t \to \infty}\bm{\xi}(t) = 0$, for all $\|\bm{\xi}(t_{0})\|<c$, uniformly in $t_{0}$; that is, for each $\bm{\epsilon}>0$, there is $t(\bm{\epsilon})>0$ such that $\|\bm{\xi}(t)\|<\bm{\epsilon}$, $\forall t \geq t_{0}+t(\bm{\epsilon})$, $\forall \|\bm{\xi}(t_{0})\|<c$.

\subsection{Model Predictive Controller Design}
\par For model simplification and computational efficiency, the vehicle dynamic model in \cite{wang2019crash} will be utilized in the process of MPC controller design for each vehicle $i$ in the platoon. For brevity, the subscripts $i$ are omitted from the dynamic model. The vehicle model with longitudinal, lateral and yaw dynamics is governed by the following equations:
\begin{equation}
\label{Eq12}
    \begin{aligned}
    m(\dot{\bar{u}}-vr)&=F_{xT}\\
    m(\dot{v}+\bar{u}r)&=F_{yf}+F_{yr}\\
    I_z\dot{r}&=F_{yf}l_f-F_{yr}l_r
    \end{aligned}
\end{equation}
where $m$ and $I_z$ are the vehicle’s mass and the moment of inertia, respectively, and $l_f$ and $l_r$ represent the distance from the vehicle's CG (center of gravity) to the front and rear axles, respectively. The vehicle’s longitudinal and lateral velocities is denote by $\bar{u}$ and $v$, respectively, $r$ is the vehicle yaw rate, $F_{xT}$ is the total longitudinal force of tires, and $F_{yf}$ and $F_{yr}$ are the total lateral forces of the front and rear tires, respectively.
\par Besides, the vehicle’s motion with respect to global coordinates is given by:
\begin{equation}
\label{Eq13}
    \begin{aligned}
    \dot{X}=\bar{u}cos\varphi-vsin\varphi,~\dot{Y}=\bar{u}sin\varphi+vcos\varphi
    \end{aligned}
\end{equation}
where $\varphi$ denotes the vehicle heading angel, $X$ and $Y$ are the vehicle longitudinal and lateral positions with respect to the global coordinate, respectively.
\par For front-wheel-steering system, the linear tire model is implemented, and the vehicle’s lateral force can be defined as:
\begin{equation}
\label{Eq14}
    \begin{aligned}
    F_{yf}&=-C_{\alpha_f}\alpha_f=C_{\alpha_f}(\delta-\frac{v+l_fr}{\bar{u}})\\
    F_{yr}&=-C_{\alpha_r}\alpha_r=C_{\alpha_r}(-\frac{v-l_rr}{\bar{u}})
    \end{aligned}
\end{equation}
where $C_{\alpha_f}$ and $C_{\alpha_r}$ denote the cornering stiffness of the front and rear tires, respectively, $\alpha_f$ and $\alpha_r$ refer to slip angle of the front and rear tires, respectively, and $\delta$ denotes the steering angle.
\par Subsequently, the state space model of a single vehicle is defined as:
\begin{equation}
\label{Eq15}
    \begin{aligned}
        &\bm{x}(t+k+1)=\bm{A}_M\bm{x}(t+k)+\bm{B}_M\bm{u}_{c}(t+k)\\
    &\bm{y}(t+k)=\bm{C}_M\bm{x}(t+k)
    \end{aligned}
\end{equation}
where the subscript $t$ denotes the current timestamp, $k$ is the prediction step, $\bm{x}=[\bar{u}~v~r~\varphi~X~Y]^T$, $\bm{u}_c=[F_{xT},\delta]^T$, $\bm{y}=[Y,\bar{u}]^T$, $\bm{A}_M$, $\bm{B}_M$, $\bm{C}_M$ denote the platoon system matrix, control matrix and output matrix, respectively. The desired outputs $Y_{des}$ and $\bar{u}_{des}$ are received from the global planning module of the platoon (assumed to be known) and UCL, respectively.
\begin{equation}
\label{Eq16}
    \begin{aligned}
    \bm{y}_{des}=[Y_{des}~\bar{u}_{des}]^T
    \end{aligned}
\end{equation}
\par Besides, the actuator capacities of the vehicle are processed as
\begin{equation}
\label{Eq17}
    \begin{aligned}
    |\delta| \leq \delta_{max},~~|F_{xT}| \leq F_{xT\_max}
    \end{aligned}
\end{equation}
where $\delta_{max}$ and $F_{xT\_max}$ denote the maximum steering angle and the maximum longitudinal force, respectively.
\par There are three types of obstacles that should be avoided on traffic roads, including non-crossable obstacles such as obstacle vehicles or pedestrians, crossable bumps, and road boundaries \cite{wang2019crash}.
\par For brevity, the functions of crossable obstacle and road boundaries are not shown here, which can be found in \cite{wang2019crash}. The potential field of non-crossable obstacles can be expressed as a function of safety distance $\bm{s}$ \cite{schulman2013finding}, which is denoted as follows:
\begin{equation}
\label{Eq19}
    \begin{aligned}
    U_{NC}(X,Y)=\frac{a}{\bm{s}^{b}}=\frac{a}{\bm{s}(\frac{dX}{X_{s}},\frac{dY}{Y_{s}})^{b}}
    \end{aligned}
\end{equation}
where $a$ and $b$ are the density and shape parameters of the non-crossable potential field, respectively, $dX$ and $dY$ are the longitudinal and lateral distances between the ego vehicle and the obstacle vehicle, respectively. The safety distance $\bm{s}$ is normalized by the safe longitudinal $X_{s}$ and lateral distance $Y_{s}$ from obstacle, which are defined as:
\begin{equation}
\label{Eq18}
    \begin{aligned}
    s&=((\frac{dX}{X_{s}}cos\bar{\theta}+\frac{dY}{Y_{s}}sin\bar{\theta})^2+(-\frac{dX}{X_{s}}sin\bar{\theta}+\frac{dY}{Y_{s}}cos\bar{\theta})^2)^{\frac{1}{2}}\\
    X_{s}&=X_0+\bar{u}T_0+\frac{\bigtriangleup\bar{u}^2_{a}}{2a_n}\\
    Y_{s}&=Y_0+(\bar{u}sin\theta+\bar{u}_{o}sin\theta)T_0+\frac{\bigtriangleup v^2_{a}}{2a_n}\\
    \end{aligned}
\end{equation}
where $X_0$ and $Y_0$ are the minimum longitudinal and lateral distances, respectively, $T_0$ is the safe time gap, $\bar{u}$ and ${\bar{u}}_{o}$ are the longitudinal velocities of the ego and obstacle vehicles, respectively, $\theta$ is the heading angle between the ego and obstacle vehicles, $\bar{\theta}$ is the heading angle between $s$ and $\frac{dX}{X_{s}}$, and ${\bigtriangleup\bar{u}}_{a}$ and $\bigtriangleup v_{a}$ are the longitudinal and lateral approaching velocities, respectively. $a_n$ is the comfortable acceleration. For brevity, the safe distance $s$ is given under the condition of non-contact from \cite{rasekhipour2016potential}, where more details can be found. 

\par Hence, the optimization problem can be expressed as follows:
\begin{equation}
\label{Eq20}
\begin{aligned}
    &\underset{ \bm{u}_{c},\varepsilon}{min}\sum\limits_{k=1}^{N_p}(\|\bm{y}(t+k,t)-\bm{y_{des}}(t+k,t)\|_{\bm{T}}^2+\|\varepsilon_k\|_{\bm{H}}^2\\
    &+\|\bm{u}_{c}(t+k-1,t)-\bm{u}_{c}(t+k-2,t)\|_{\bm{R}}^2\\
    &+\|\bm{u}_{c}(t+k-1,t)\|_{\bm{S}}^2+\|U_{NC}(t+k,t)\|_{\bm{Q_{obs}}})\\
    &s.t.(k=1,\dots,N_p)\\
    &\bm{x}(t+k+1)=\bm{A}_M\bm{x}(t+k)+\bm{B}_M\bm{u_c}(t+k)\\
    &\bm{y}(t+k)=\bm{C}_M\bm{x}(t+k)\\
    &\bm{u}_{cmin}(t+k-1)\leq\bm{u}_c(t+k-1)\leq\bm{u}_{cmax}(t+k-1)\\
    &\Delta\bm{u}_{cmin}(t+k-1)\leq\Delta\bm{u}_c(t+k-1)\leq\Delta\bm{u}_{cmax}(t+k-1)\\
    &\bm{u}_c(t+k)=\bm{u}_c(t+k-1), \forall k\geq N_c
\end{aligned}
\end{equation}
where $N_p$ and $N_c$ are the prediction and control horizons, respectively, $\varepsilon_k$ is the vector of slack variables at k steps ahead of the current
time, $\bm{x}(t+k)$ represents the predicted state values of the system, $\bm{y}(t+k)$ denotes the predicted outputs of the system over the prediction horizon, $\bm{u}_{cmin}$ and $\bm{u}_{cmax}$ are the lower and upper bounds of the actuator, respectively, and $\Delta\bm{u}_{cmin}$ and $\Delta\bm{u}_{cmax}$ are the variation ranges of control variables at each time. In the cost function shown in (\ref{Eq20}), matrices $\bm{T}$, $\bm{H}$, $\bm{R}$, $\bm{S}$ and $\bm{Q}_{obs}$ are the weight matrices corresponding to the tracking of the desired output, slack variables, the variation of the control input, the control input, and the APF, respectively. 
\par Due to the non-linearity and non-convexity of APF, it needs to be transformed to corresponding quadratic convex optimization problem, as shown in \cite{wang2019crash, rasekhipour2018autonomous}. After the convex process, the optimal control problem is a convex quadratic optimization problem. It is similar to a corresponding nonlinear problem solved by Sequential Quadratic Programming (SQP) in one sequence. Based on the upper bound method from \cite{boggs1995sequential}, the closer the problem's initial value is to the minimum, which is equivalent to the anticipated vehicle point being closer to the vehicle position at the minimum, the smaller the optimization error. The performance of this solver will be presented in the case studies.

\subsection{Framework Design}
\begin{figure*}[t]
    \centering
    \includegraphics[width=16cm]{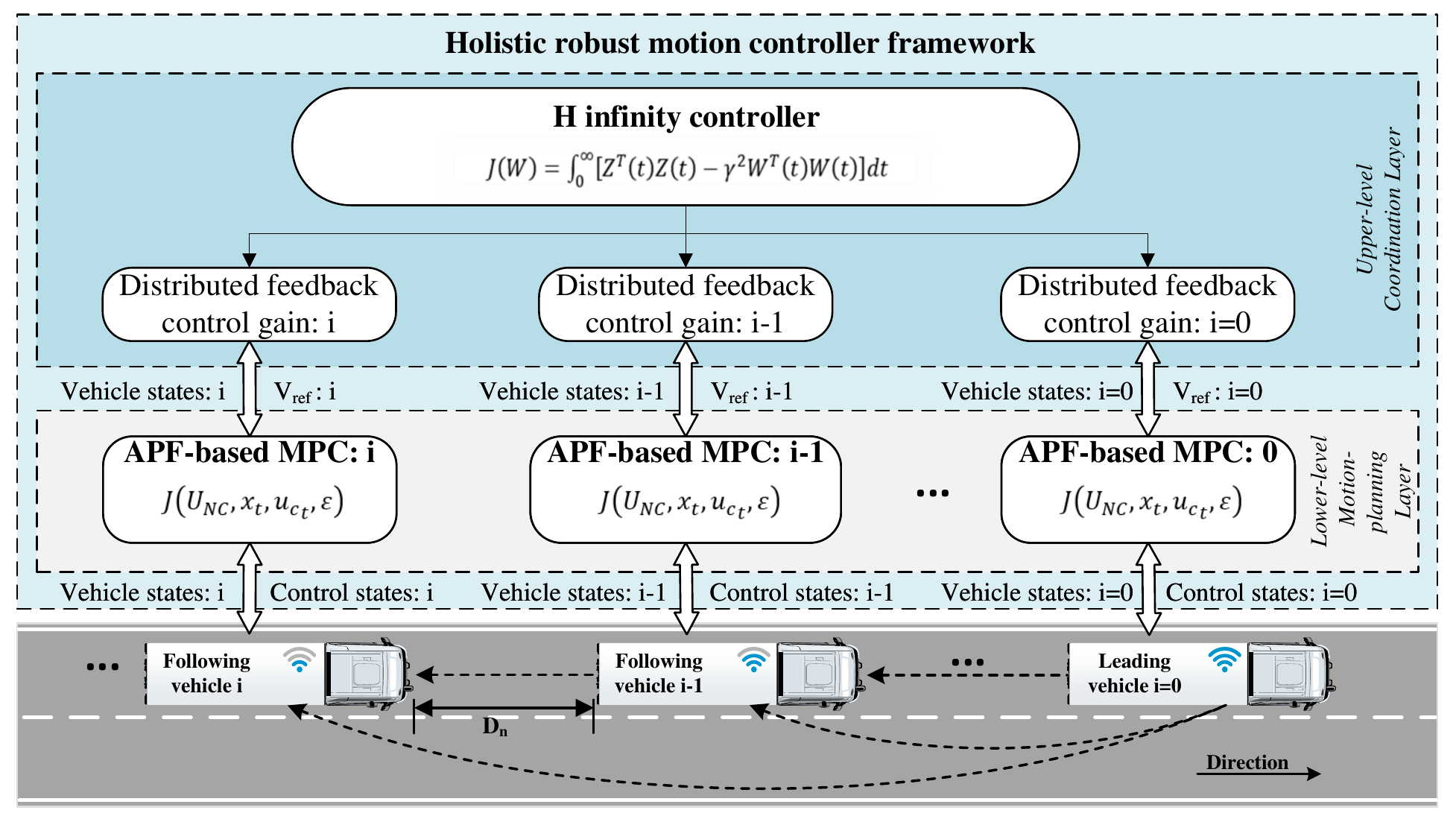}
    \caption{Overall structure of the holistic robust motion controller framework for autonomous platooning.}
    \label{Fig2}
\end{figure*}
Fig. \ref{Fig2} depicts the overall structure of the holistic robust motion controller framework. In this paper, the MCF divides the platooning control into upper-level coordination layer (UCL) and lower-level motion-planning layer (LML) to solve the comprehensive platooning control problem. Due to complex individual movements of vehicles in autonomous platoon, the control of autonomous platoon's longitudinal string stability should be considered from a systematic perspective. As depicted in Fig. \ref{Fig2}, the proposed H-infinity feedback controller is implemented in the UCL to achieves the longitudinal coordination and robustness of the platoon from a systematic perspective. On the other hand, the distributed controllers should be deployed to ensure functional independence of lateral control for the flexibility and safety of the platoon system. The proposed APF-based MPC is implemented in the LML, which guarantees the independence of the lateral control. And it achieves multi-objective control based on the reference signal from the UCL, multiple constraints of vehicle dynamics and the APF. 
\par With the UCL and LML connected, the proposed MCF achieves system robustness and a good balance between longitudinal string stability and lateral safety under the help of robust control and multi-objective rolling optimization. When the autonomous platoon is proceeding without encountering obstacles, the proposed MCF mainly conducts the commands from H infinity controller, which is consistent with Theorem 1 so that it can maintain string stability of the system. When there are obstacles vehicles intruding, the proposed MCF can conduct obstacle avoidance based on APF to guarantee the safety of vehicles in the platoon. Moreover, a hierarchical structure achieves a more comprehensive performance without complicating the controller synthesis compared with a consensus method. And it ensures the scalability of the algorithm for other safety-related issues. For instance, the UCL of the proposed MCF can be modified with other specialized robust controllers.
\par For simplicity, the platoon constructed for H-infinity controller design consists of homogeneous vehicles. It is assumed that H-infinity controller is designed offline according to the vehicle number and parameter variation of the platoon. And the corresponding feedback control gains are stored in distributed storage devices of each vehicle in the platoon, which can be searched online and obtained by each vehicle in the platoon when the platoon is initially organized. Furthermore, it is assumed in this study that the perception and global planning modules are functioning properly, and the data processed by these two modules are accurate and ready to use. In other words, this study focuses on the motion planning of the autonomous platoon.
\par Besides, the LML tends to choose aggressive maneuvers in order to follow the reference signal delivered from the UCL, while the platoon is avoiding the obstacles. Thus, for a smoother and safer longitudinal and lateral collaborative controller performance, an adaptive control algorithm for adjusting the control weight of the reference signal is designed based on the APF value. The algorithm enables the control weight of the reference signal to be reduced while following the obstacle avoidance procedure. The weight matrix $\bm{T}$ includes two parts, i.e., $\bm{T}=[Q_{Y},~Q_{ref}]$, where $Q_{Y}$ and $Q_{ref}$ are weight parameters for $Y_{des}$ and $\bar{u}_{des}$, respectively. To achieve the goal aforementioned, the  $Q_{ref}$ is designed as follows:
\begin{equation}
\label{Eq21}
    \begin{aligned}
    Q_{ref}=Q_0(\frac{1}{\sigma_0}+\sum\limits_{j=1}^m\frac{1}{\sigma_j+U_{NC_j}})
    \end{aligned}
\end{equation}
where $Q_{ref}$, $Q_0$, $\sigma_j$ represent the control weight of reference signal, the control weight of reference signal while proceeding the regular platooning, and the adjustment parameter of the $j^{th}$ obstacle, respectively. The pseudocode of the working procedure for the proposed MCF is demonstrated in Table I.
\begin{table}
    \caption{Pseudocode of the working procedure for the holistic robust motion controller framework}
    \centering
    \begin{tabular}{l}
    \toprule
    \toprule
    \textbf{Algorithm 1} Pseudocode of the working procedure \\
    \hline
    ~1: \textbf{Input:}\\
    ~2:  ~~~~~The number of vehicles in the platoon $N$, desired distance\\
    ~~~~~~~~~between each vehicle $D_n$;\\
    ~3: ~~~~~The state of obstacle vehicles: $\bm{V}_{obs}$;\\
    ~4: ~~~~~The state of vehicles in the platoon:$\bm{x}_{pla}$;\\
    ~5: \textbf{Output:}
    The longitudinal force $F_{x,i}$ and the steering angle $\delta_{f,i}$;\\
    ~6: \textbf{Initialization:}\\
    ~7: ~~~~~Search for the H-infinity feedback control gain $\bm{K}$ of UCL;\\
    ~8: ~~~~~Set the weight matrices $\bm{T}$, $\bm{Q_{obs}}$, $\bm{S}$, $\bm{H}$, $\bm{R}$ of LMP;\\
    ~9: \textbf{While platoon is proceeding:}\\
    10: ~~~~~ Receive the state of obstacle vehicles: \\
    ~~~~~~~~~~$\bm{V}_{obs}=\{X_{s,j},Y_{s,j}, v_{xs,j},v_{ys,j}\}_{j=1:M}$ \# $M$ is the number \\ 
    ~~~~~~~~~~of obstacle vehicles;\\
    11:~~~~~~Receive the state of vehicles in the platoon:\\
    ~~~~~~~~~~$\bm{x}_{pla}=\{\bar{u}_i,v_i,r_i,\varphi_i, X_i,Y_i\}_{i=1:N}$;\\
    12: ~~~~~UCL calculates the reference velocity for each vehicle $i$\\
    ~~~~~~~~~~in the platoon: ${V}_{ref,i}$;\\
    13: ~~~~~LML calculates the $APF$ value for each vehicle $j$:~$U_{NC_j}$;\\
    14: ~~~~~LML calculates the adaptive weight $Q_{ref,i}$ for each vehicle \\ ~~~~~~~~~~$i$ to update $\bm{T}$;\\
    15:~~~~~~\textbf{For}~~$i=1:N$~~\textbf{do:}\\
    16:~~~~~~~~~~$J_i=0$ \# The cost of LML for each vehicle $i$ in the platoon\\
    17:~~~~~~~~~~\textbf{For}~~$k=1:N_p$~~\textbf{do:} \# $N_p$ is the prediction horizon in LML\\
    18:~~~~~~~~~~~~~Predict the states of obstacle vehicles and ego vehicle to\\
    ~~~~~~~~~~~~~~~~~obtain $U_{NC_j}(t+k),\bm{y}_i(t+k)$, $\bm{y}_{\bm{des}_i}(t+k)$,\\
    ~~~~~~~~~~~~~~~~~$\bm{u}_{c_i}(t+k-1)$;\\
    19:~~~~~~~~~~~~~~\# Sum the loss at each prediction step:\\
    20:~~~~~~~~~~~~~~$J_i=J_i+\|\bm{y}_i(t+k,t)-\bm{y}_{\bm{des}_i}(t+k,t)\|_{\bm{T}}^2$\\
    ~~~~~~~~~~~~~~~~~$+\|U_{NC_j}(t+k,t)\|_{\bm{Q_{obs}}}+\|\bm{u}_{c_i}(t+k-1,t)\|_{\bm{S}}^2+$\\
    ~~~~~~~~~~~~~~~~~$\|\varepsilon_{k_i}\|_{\bm{H}}^2+\|\bm{u}_{c_i}(t+k-1,t)-\bm{u}_{c_i}(t+k-2,t)\|_{\bm{R}}^2$\\
    21:~~~~~~~~~~\textbf{End}\\
    22:~~~~~~~~~~Minimize $J_i$ and outputs the control states: the longitudinal\\
    ~~~~~~~~~~~~~force $F_{x,i}$ and the steering angle $\delta_{f,i}$ to control the vehicle $i$.\\
    23:~~~~~~\textbf{End}\\
    \toprule
    \toprule
    \end{tabular}
    \label{Tab1}
\end{table}

\section{Case Studies}
In this section, the functionality of the proposed MCF is verified by designing and analyzing three critical scenarios. The implementation of the scenarios is based on the co-simulation of MATLAB/Simulink and TruckSim. The transportation truck is modeled in TruckSim with great fidelity. Table II shows part of the simulation parameters.
\begin{table}[t]
    \caption{Vehicle and simulation parameters used in case studies}
    \centering
    \begin{tabular}{cccc}
    \toprule
    \toprule
    \textbf{Parameters} & \textbf{Value} & \textbf{Parameters} & \textbf{Value} \\
    \hline
    $m$ & 5760 $kg$ & $N_c$ & 5 \\
    $l_f$ & 1.11 $m$ &$[Q_{Y},~Q_{0}]$& [300 375] \\
    $l_r$ & 3.89 $m$ & $\bm{S}$ & [0.001 1] \\
    $l_z$ & 34802.6 $kg \cdot m^2$ & $\bm{R}$ & [100 2000]\\
    $C_{\alpha_f}$ & 259752 $N/rad$ & $\bar{K}_1$ & [2.156 3.175 0.998] \\
    $C_{\alpha_r}$ & 260000 $N/rad$ & $\bar{K}_2$ & [0.306 0.239 0.065]\\
    $h_1$ & 250 ms & $N_p$ & 25 \\
    $a$ & 10000 & $\sigma_0$ & 2 \\
    $b$ & 1.23 & $\sigma_j$ & 4\\
    \toprule
    \toprule
    \end{tabular}
    \label{Tab2}
\end{table}

\begin{figure}
    \vspace{-0.2cm}
    \centering
    \includegraphics[width=9cm]{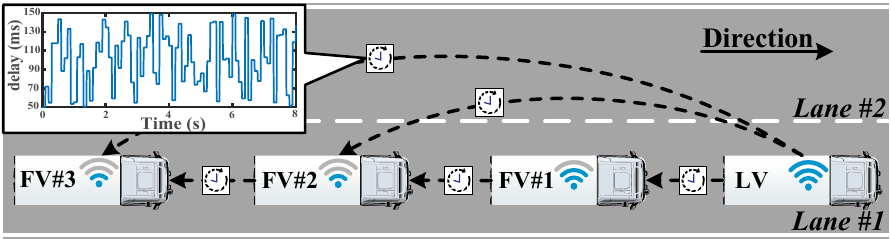}
    \caption{Platooning scenario under time-varying communication delay.}
    \vspace{-0.2cm}
    \label{Fig3}
\end{figure}

\subsection{Platooning Scenario under time-varying communication delay}
\begin{figure}
    \vspace{-0.3cm}
    \centering
    \includegraphics[width=8.5cm]{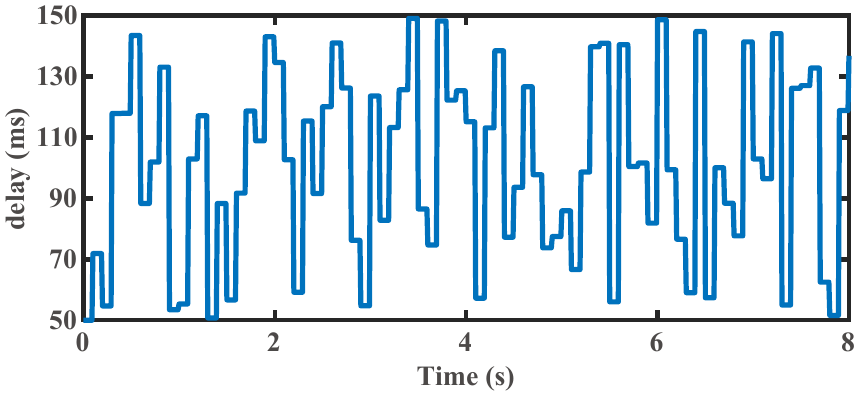}
    \caption{The time-varying delay in V2V communication.}
    \vspace{-0.28cm}
    \label{Fig4}
\end{figure}

\begin{figure}
    \vspace{-0.5cm}
    \centering
    \subfigure[]{
    \includegraphics[width=8.5cm]{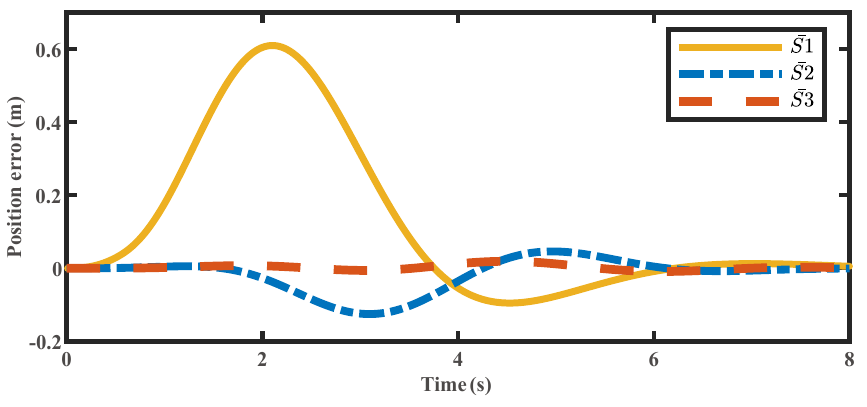}
    \label{Fig5a}
    }
    \subfigure[]{
    \includegraphics[width=8.5cm]{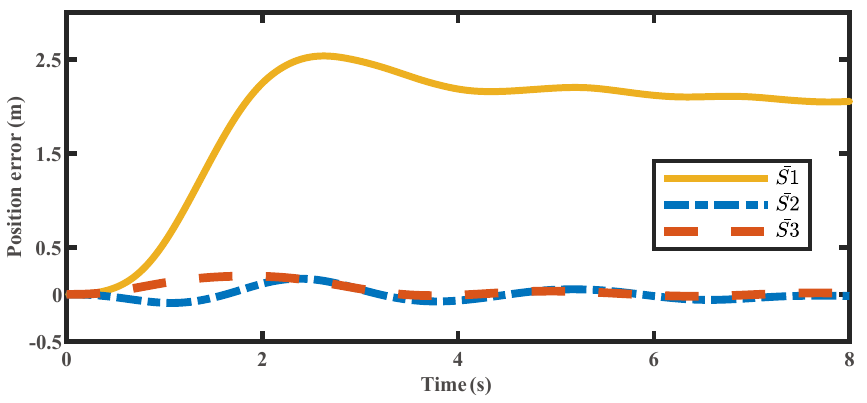}
    \label{Fig5b}
    }
    \caption{Position error: (a) without delay and (b) with time-varying delay.}
    \label{Fig5}
    \vspace{-0.2cm}
\end{figure}
\par In this scenario, the string stability of platoon utilizing the proposed MCF under time-varying delay is demonstrated. As depicted in Fig. \ref{Fig3}, the platoon consists of four commercial vehicles, and the inter-vehicle spacing is set to 20m. The initial longitudinal velocity of the vehicles in the platoon is 20 m/s, and the leading vehicle will accelerate from 20 m/s to 22 m/s with an overshoot, as in \cite{zheng2016distributed}. The time-varying communication delay is bounded between 50-150 ms and its derivative is less than one, as shown in Fig. \ref{Fig4}. The time-varying communication delays in different channels are given with different random seeds to achieve variation.
\par The platoon position error without delay and with delay are shown in Fig. \ref{Fig5a} and Fig. \ref{Fig5b}, respectively. As indicated, the proposed MCF is eligible to maintain the platoon longitudinal string stability with and without time-varying communication delay. The  position error converges to the equilibria  after $t_{0}=6s$, thus eventually enables the position error to converge to the asymptotical stability, which is consistent with \textbf{\textit{Theorem 1}}. Furthermore, the convergence of position error is uniform, and its deviation is independent of $t_{0}$. It is complied with the definition of uniformly asymptotically stable in \textbf{\textit{Lemma 1}}. 
\par
\begin{figure}
    \vspace{-0.2cm}
    \centering
    \subfigure[]{
    \includegraphics[width=8.5cm]{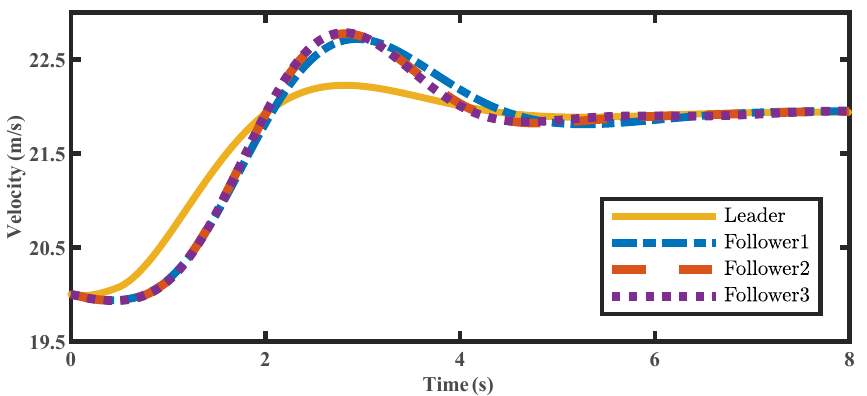}
    \label{Fig6a}
    }
    \subfigure[]{
    \includegraphics[width=8.5cm]{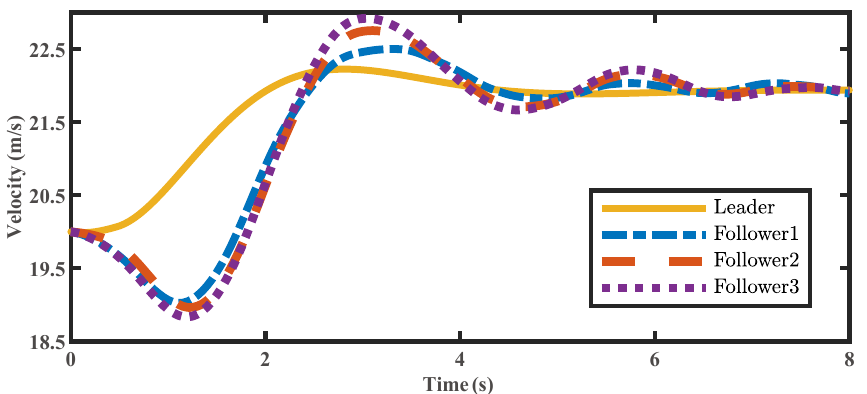}
    \label{Fig6b}
    }
    \caption{Velocity: (a) without delay and (b) with time-varying delay.}
    \label{Fig6}
    \vspace{-0.2cm}
\end{figure}

 As shown in the Fig. \ref{Fig5a}, when the platoon proceeds without the communication delay, the position error of each vehicle in the platoon is less than 1 m and converges to 0 after 6s. Fig. \ref{Fig5b} shows that when the platoon proceeds under the time-varying communication delay, there is a slight position error deviation after 4s between the leading vehicle and each following vehicle, because the transmitted vehicular states of the leading vehicle is delayed \cite{de2019stability}. However, the position error between each following vehicle can be controlled effectively and converges to 0. Fig. \ref{Fig6a} and Fig. \ref{Fig6b} show the velocity of vehicles in the platoon without and with the time-varying delay, respectively. As shown in Fig.\ref{Fig6}, when overshoots occur between 2s and 4s, the proposed MCF enables the following vehicles to track the velocity of the leading vehicle with or without the communication delay and converge to the uniform asymptotic stability after 6s.
\vspace{-0.2cm}
\subsection{Merging Scenario}
\begin{figure}
    \centering
    \includegraphics[width=9cm]{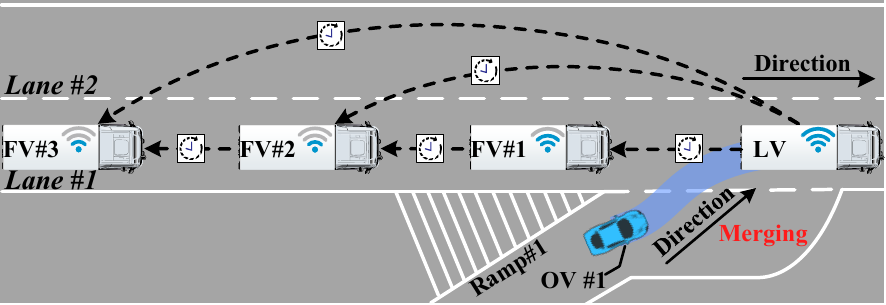}
    \caption{Merging scenario with an obstacle vehicle splitting platoon under time-varying communication delay.}
    \vspace{-0.5cm}
    \label{Fig7}
\end{figure}
In this scenario, an obstacle vehicle on the ramp aggressively merges into the platoon cruising lane, and the platoon is cut in by the sudden-appearing obstacle, as depicted in Fig. \ref{Fig7}. Specifically, this scenario consists of four commercial vehicles within the platoon and an obstacle vehicle. And the obstacle vehicle on the ramp merges into the first lane with a velocity of 15 m/s. The communication delay is bounded between 15-150 ms. Under the time-varying communication delay, the platoon is initialized at a velocity of 20 m/s and a inter-vehicle space of 50 m on the first lane. 
\begin{figure}
    \centering
    \includegraphics[width=9cm,height=4.5cm]{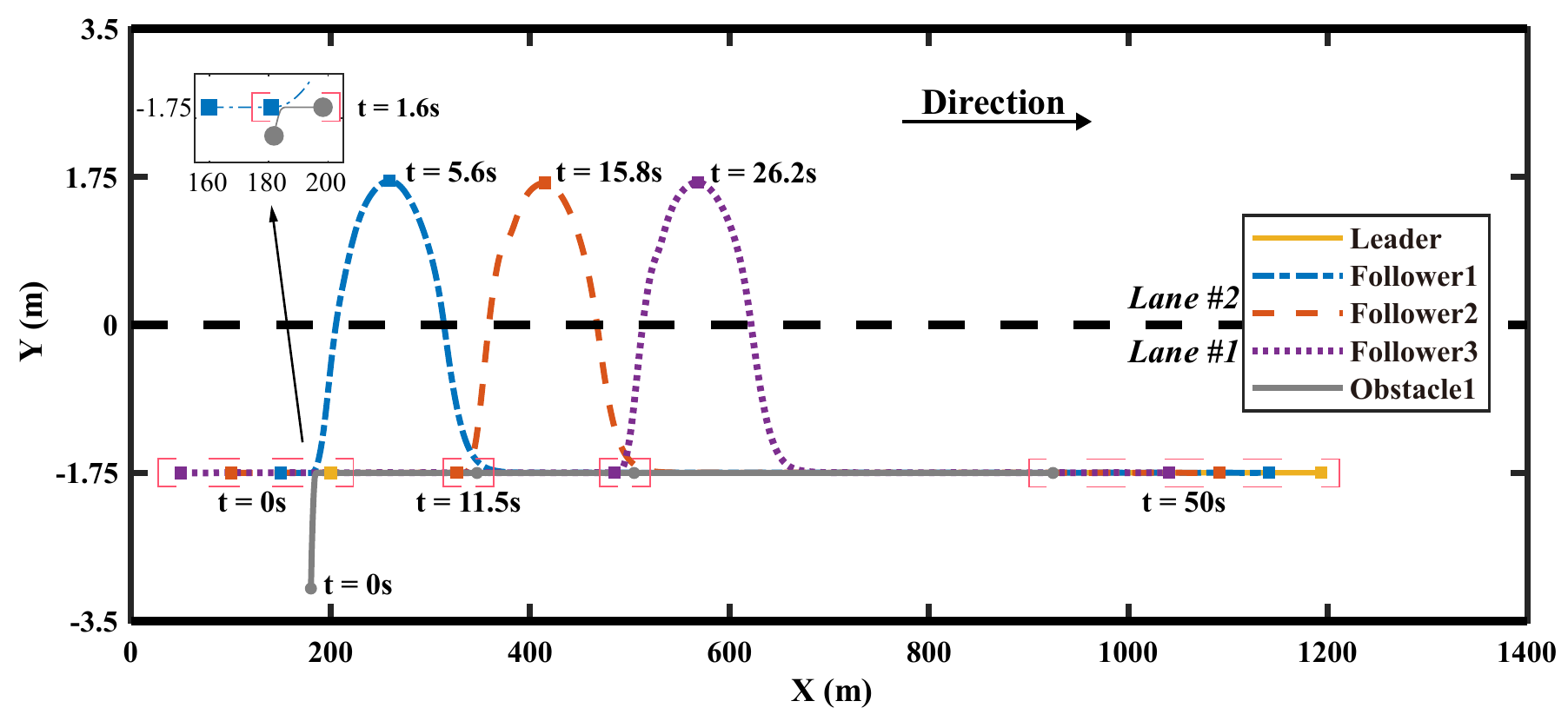}
    \caption{Movement of the platoon in merging scenario.}
    \vspace{-0.5cm}
    \label{Fig8}
\end{figure}
\begin{figure}
    \centering
    \includegraphics[width=8.5cm]{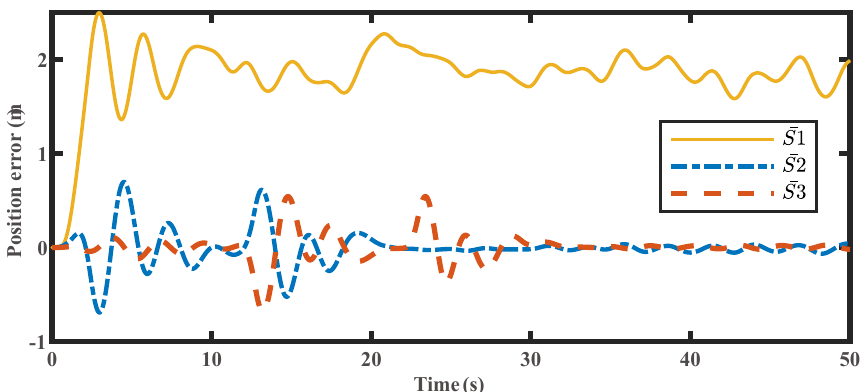}
    \caption{Position error of the platoon in merging scenario.}
    \vspace{-0.5cm}
    \label{Fig9}
\end{figure}
\begin{figure}
    \vspace{-0.5cm}
    \centering
    \includegraphics[width=8.5cm]{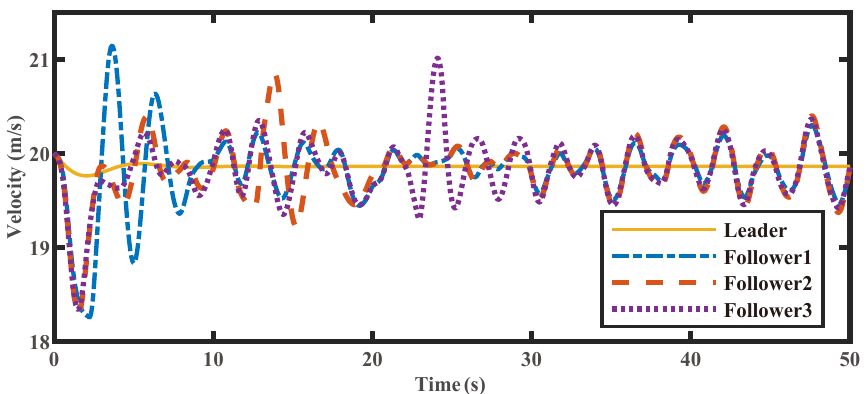}
    \caption{Velocity of the platoon in merging scenario.}
    \vspace{-0.5cm}
    \label{Fig10}
\end{figure}
\par Fig. \ref{Fig8} shows the movement diagram of the platoon. As it is depicted, the obstacle vehicle merges into the first lane from the ramp on the right side and blocks the following vehicles. Starting from 1.6s, the following vehicle 1 is the first vehicle that encounters the obstacle vehicle. It starts to overtake the obstacle vehicle due to the slower velocity of the obstacle vehicle. After encountering the obstacle vehicle, the remaining the following vehicles in the platoon conduct the obstacle avoidance motion planning to catch up with the preceding vehicles. Eventually, the platoon realizes the safety operations under the merging scenario. 

\par The position error is shown in Fig. \ref{Fig9}. Under the time-varying communication delay, a position error deviation exists between each following vehicle and the leading vehicle. Moreover, due to the obstacle avoidance, the position error fluctuates but finally converges to uniform asymptotic stability after 30s. The velocity diagram shown in Fig. \ref{Fig10} demonstrates that the velocity converge to the uniform asymptotic stability after each vehicle in the platoon accomplishes the obstacle avoidance.

\subsection{Obstacle Avoidance Scenario}

In this part, a critical scenario is demonstrated where the platoon is blocked by obstacle vehicles on the road. Specifically, this scenario consists of four commercial vehicles within the platoon and two obstacle vehicles. The communication delay is bounded between 15-150 ms. Under the time-varying communication delay, the platoon is initialized at a velocity of 18 m/s and a inter-vehicle space of 50 m on the first lane. The obstacle vehicle 1 is initialized with a velocity of 15 m/s ahead of the platoon. While the platoon is avoiding the obstacle vehicle 1, the obstacle vehicle 2 will appear with a velocity of 17 m/s on the second lane and block the following vehicles’ obstacle avoidance for a certain period of time, as shown in Fig. 11.
\begin{figure}
    \centering
    \includegraphics[width=9cm]{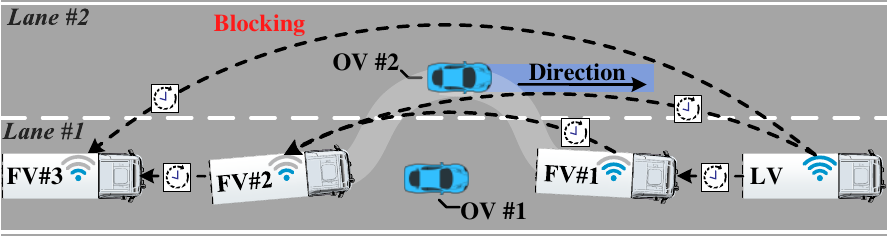}
    \caption{Obstacle avoidance scenario with critical blocking moment under time-varying communication delay.}
    \vspace{-0.3cm}
    \label{Fig11}
\end{figure}
\par
\begin{figure}
    \vspace{-0.5cm}
    \centering
    \includegraphics[width=9cm,height=4.5cm]{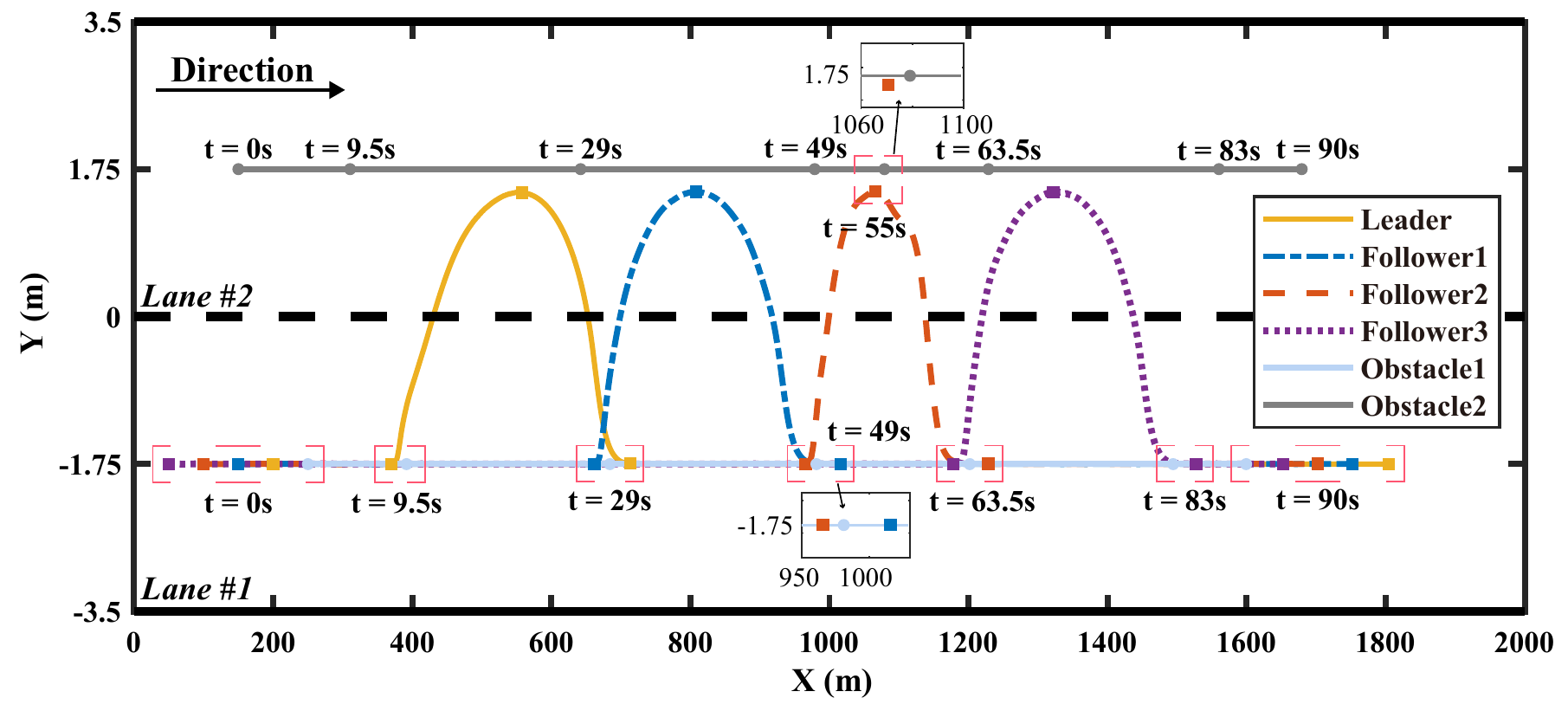}
    \caption{Movement of the platoon in obstacle avoidance scenario.}
    \vspace{-0.3cm}
    \label{Fig12}
\end{figure}

 The platoon movement diagram is shown in Fig. \ref{Fig12}. When each vehicle in the platoon encounters obstacle vehicle 1, it conducts the obstacle avoidance asynchronously to reduce the lane space requirement and risk. At 49s, the obstacle vehicles temporarily block the road. Due to the blocking, the following vehicle 2 needs to balance the trade-off between driving safety and string stability of the platoon, which makes its path less moderate than the others. With the proposed MCF, all the vehicles within the platoon safely carry out lane changing and overtaking. 

\begin{figure}
    \centering
    \includegraphics[width=8.5cm]{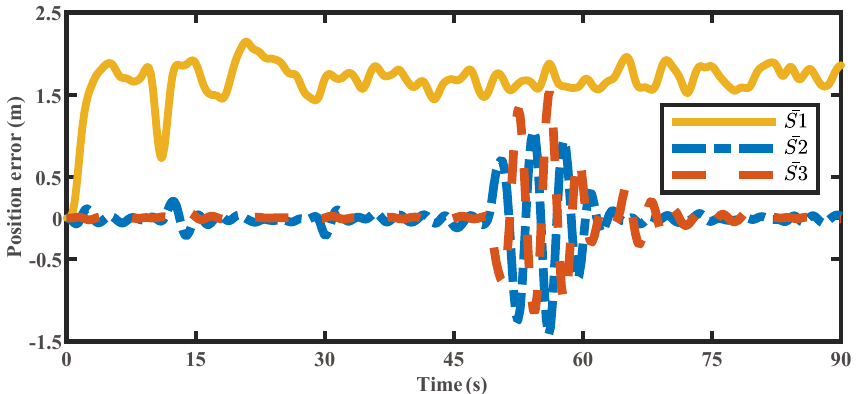}
    \caption{Position error of the platoon in obstacle avoidance scenario.}
    \vspace{-0.2cm}
    \label{Fig13}
\end{figure}

\begin{figure}
    \centering
    \includegraphics[width=8.5cm]{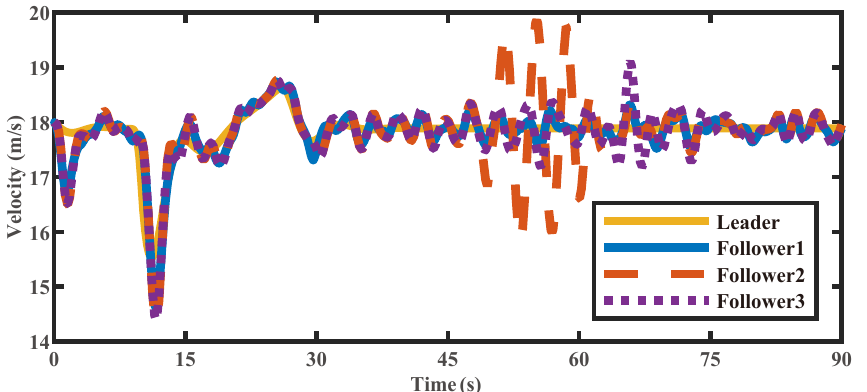}
    \caption{\textbf{Velocity of the platoon in obstacle avoidance scenario.} The fluctuation of Follower 2 at 49s is due to trade-off between the driving safety and string stability.}
    \vspace{-0.2cm}
    \label{Fig14}
\end{figure}
\begin{figure}
    \centering
    \includegraphics[width=8.5cm]{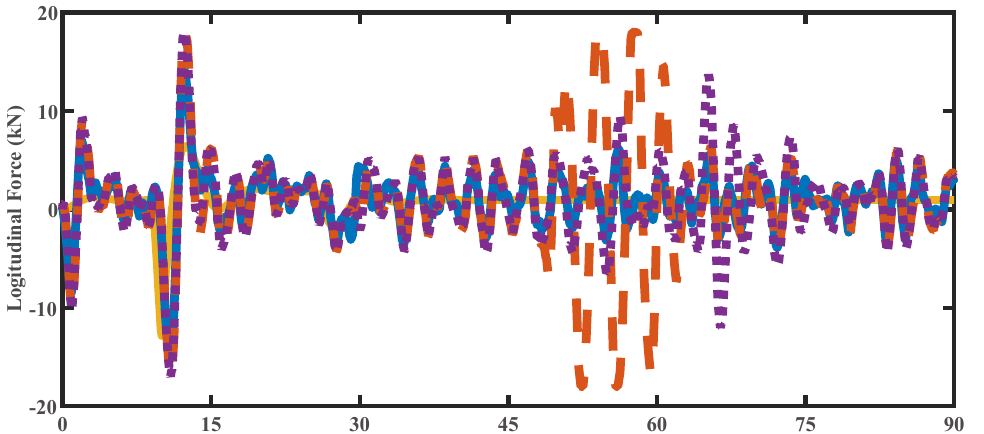}
    \includegraphics[width=8.5cm]{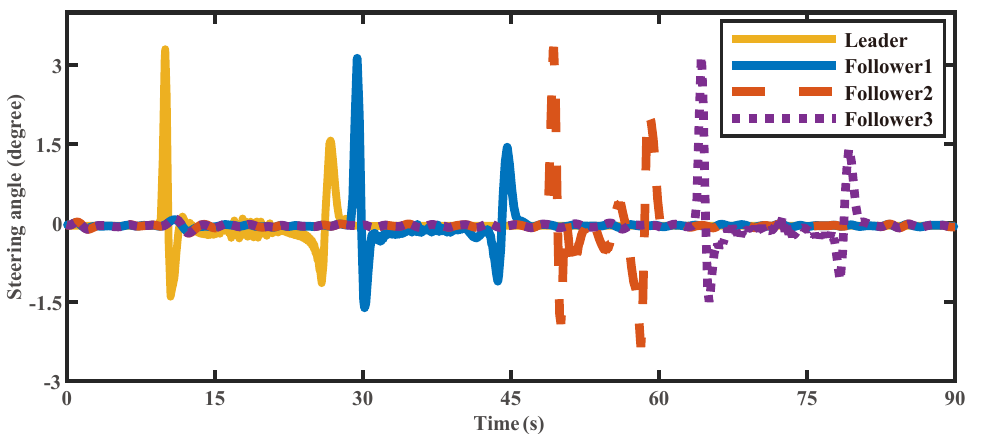}
    \caption{Longitudinal force and steering angle commands in obstacle avoidance scenario.}
    \vspace{-0.2cm}
    \label{Fig15}
\end{figure}
\begin{figure}
    \centering
    \includegraphics[width=8.5cm]{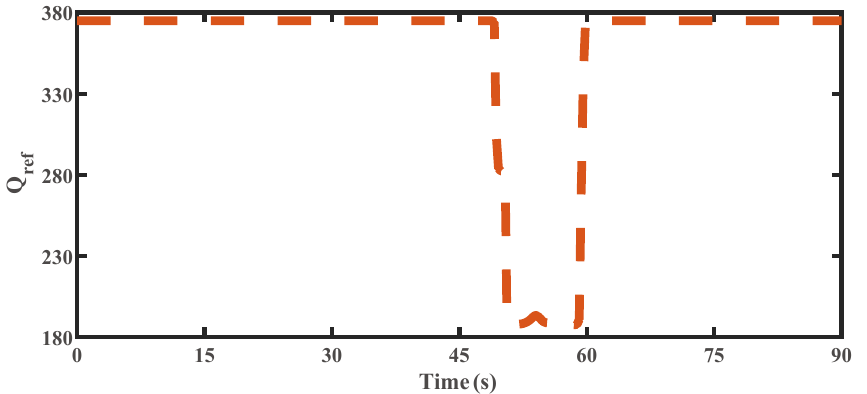}
    \caption{$Q_{ref}$ of following vehicle 2 in obstacle avoidance scenario.}
    \vspace{-0.2cm}
    \label{Fig16}
\end{figure}
\par Fig. \ref{Fig13} depicts the position error. It can be noted that the position error fluctuates during obstacle avoidance but eventually converges to uniform asymptotic stability. Due to the existence of communication delay, there is a position error deviation between each following vehicles and the leading vehicle. The velocity diagram of the platoon is shown in Fig. \ref{Fig14}. And the control commands is shown in Fig. 15. When each vehicle in the platoon encounters the obstacle vehicle 1, it first decelerates to keep safe distance, and then accelerates to change the lane. During the blocking, there is a fluctuation of the velocity of following vehicle 2. Due to the increase in the APF value, the control weight $Q_{ref}$ of the reference signal generated from the UCL for following vehicle 2 will decrease, as shown in Fig. 16. Hence, the following vehicle 2 will be controlled mainly based on the safety considerations and executing the obstacle avoidance. As the velocity of obstacle vehicle 2 is higher than that of obstacle vehicle 1, the following vehicle 2 accelerates to change to the second lane, then decelerates to follow obstacle vehicle 2. It eventually accelerates to catch up with the preceding vehicles and change back to the first lane when it is safe. After the whole platoon accomplishes obstacle avoidance, the velocity of each following vehicle converges to uniform asymptotic stability after 75s, as shown in Fig. \ref{Fig14}. This experiment demonstrates that each vehicle in the platoon can independently make safe decisions with the proposed holistic robust MCF.

\subsection{Comparison}
\par To demonstrate advantages of the proposed design over the existing ones, Table \ref{Tab3} compares the performance of the proposed MCF and the robust distributed model predictive controllers (DMPC) from \cite{zheng2016distributed} in the same platooning scenario.  $\|\bar{s}_i\|$ denotes the $\infty$-norm of the position error between the $i^{th}$ and the $(i-1)^{th}$ vehicles. As shown in Table \ref{Tab3}, the position error decreases from ${\bar{s}}_1$ to ${\bar{s}}_3$ at a faster rate along the string with the proposed MCF compared with the robust DMPC (PF or PLF). It indicates that the proposed MCF can offer a better suppression of the disturbance propagation. The maximum position error of the proposed MCF is higher than those of the robust DMPC simulated with passenger vehicle models, due to parameter difference between commercial and passenger vehicles, such as the vehicle mass. However, the maximum position error is less than 1 m with both the proposed MCF and the robust DMPC. It demonstrates the robustness of the autonomous platoon using the proposed MCF against the leader disturbance.

\begin{figure}
    \centering
    \includegraphics[width=8cm]{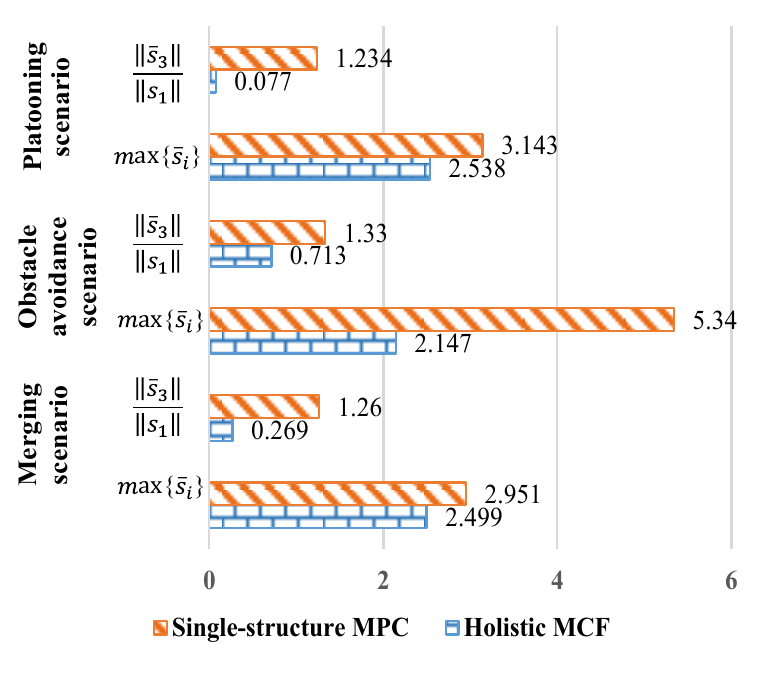}
    \caption{\textbf{Performance comparison of the holistic MCF and single-structure MPC.} The single-structure MPC is developed utilizing the APF-based MPC without considering robustness of the whole platoon.}
    \label{Fig17}
\end{figure}

To the best of our knowledge, the lateral control for autonomous platoon in critical scenarios has not been explored in literature yet. Therefore, the APF-based MPC is utilized to develop the single-structure MPC for autonomous platoon in further comparison. While the single-structure MPC doesn't consider the robustness of the autonomous platoon against time-varying communication delay, it can achieve both the longitudinal following and lateral control under an predecessor-leader following topology. For various scenarios under time-varying communication delay, the performance comparison of the holistic MCF and single-structure MPC is given in Fig. \ref{Fig17}. In platooning scenario, the holistic MCF can effectively suppress the position error from ${\bar{s}}_1$ to ${\bar{s}}_3$ under the velocity fluctuation of the leading vehicle. And it achieves a $19.2\%$ improvement in the maximum position error $max\{{\bar{s}}_i\}$ compared with single-structure MPC, as shown in Fig. \ref{Fig17}. In obstacle avoidance and merging scenarios, the holistic MCF can guarantee a better balance of the trade-off between the driving safety and string stability than single-structure MPC, with a faster decreasing rate of position error. The maximum position errors of holistic MCF are improved by $59.8\%$ and $15.3\%$ in obstacle avoidance and merging scenarios, respectively.
\begin{table}[t]
    \caption{Performance comparison}
    \centering
    \begin{tabular}{p{30pt}<{\centering}p{50pt}<{\centering}p{60pt}<{\centering}p{60pt}<{\centering}}
    \toprule
    \toprule
    \textbf{Index} & \textbf{MCF(PLF)} & \textbf{Robust DMPC(PF)}$^{*}$ & \textbf{\textbf{Robust} DMPC(PLF)}$^{*}$ \\
    \hline
    $\frac{\|\bar{s}_2\|}{\|\bar{s}_1\|}$ & 0.21 & 0.90 & 0.89 \\
    $\frac{\|\bar{s}_3\|}{\|\bar{s}_2\|}$ & 0.15  & 0.94 & 0.98 \\
    $max\{\bar{s}_i\}$ & 0.61 $m$ & 0.2$m$ & 0.2 $m$ \\
    \toprule
    \toprule
    \end{tabular}
    \footnotesize{$*$The results of the Robust DMPC(PF) and DMPC(PLF) are taken from [15] without time-varying communication delay}
    \label{Tab3}
    \vspace{-0.3cm}
\end{table}

\section{Experimental verification}
\begin{figure}
    \centering
    \includegraphics[width=8cm]{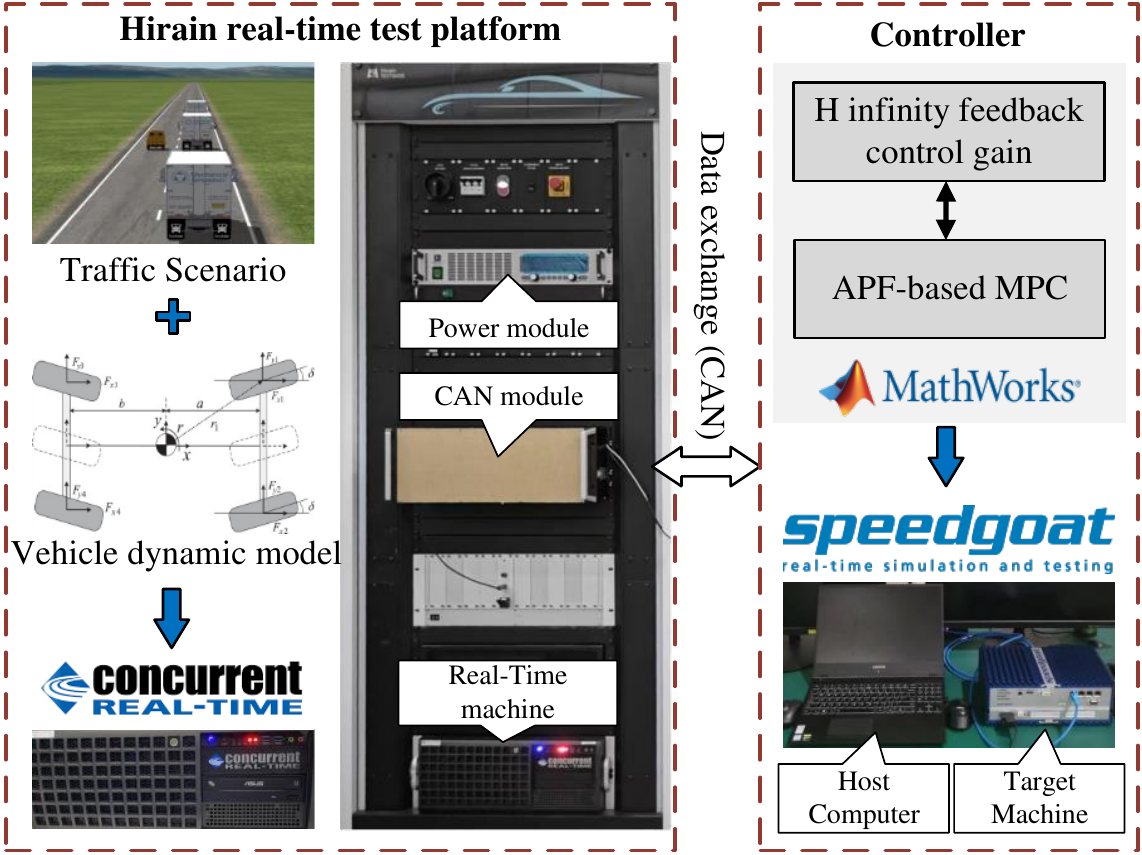}
    \caption{Hardware-in-the-loop platform}
    \label{Fig18}
\end{figure}
\begin{figure}
    \vspace{-0.5cm}
    \centering
    \includegraphics[width=9cm,height=4.5cm]{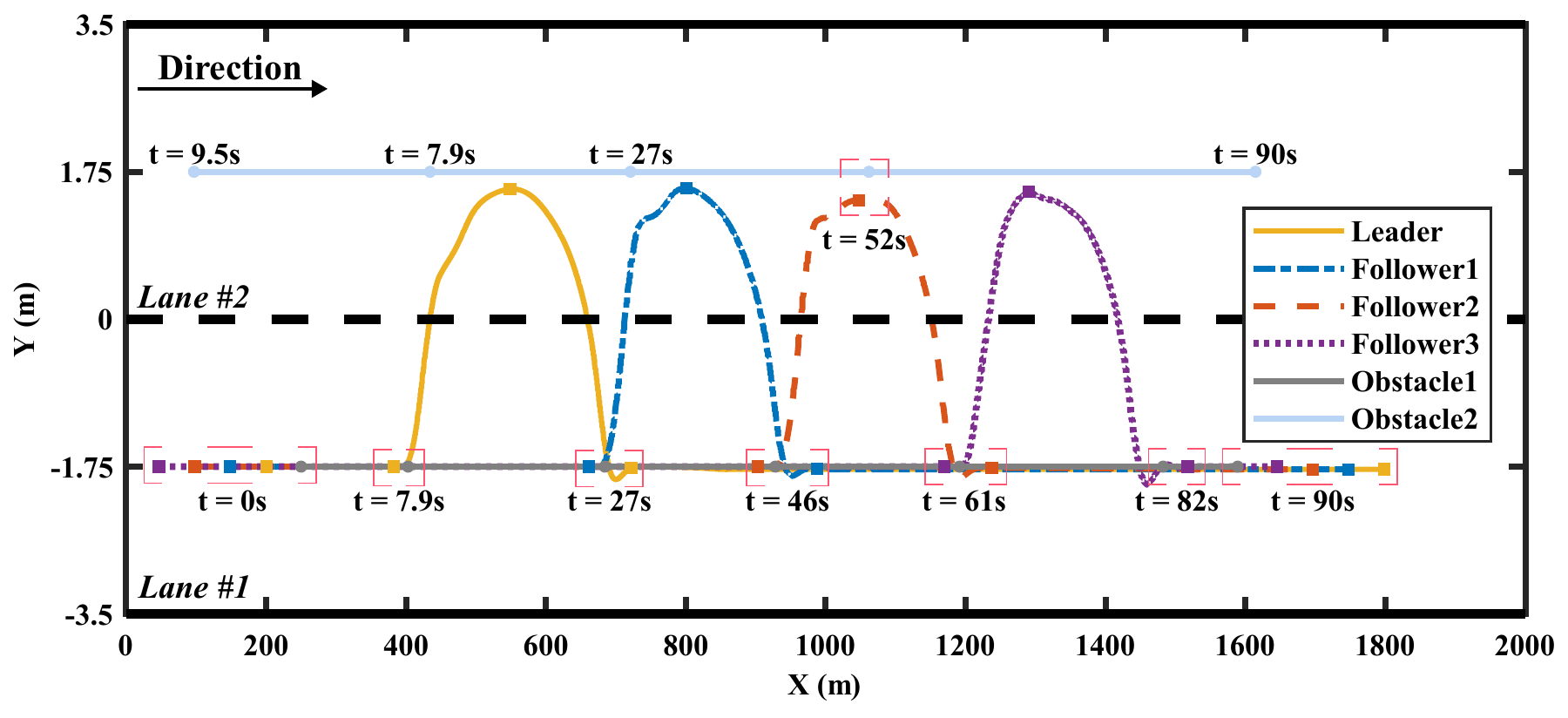}
    \caption{Movement of the platoon in HIL experiment.}
    \vspace{-0.3cm}
    \label{Fig19}
\end{figure}

\begin{figure}
    \centering
    \includegraphics[width=8cm]{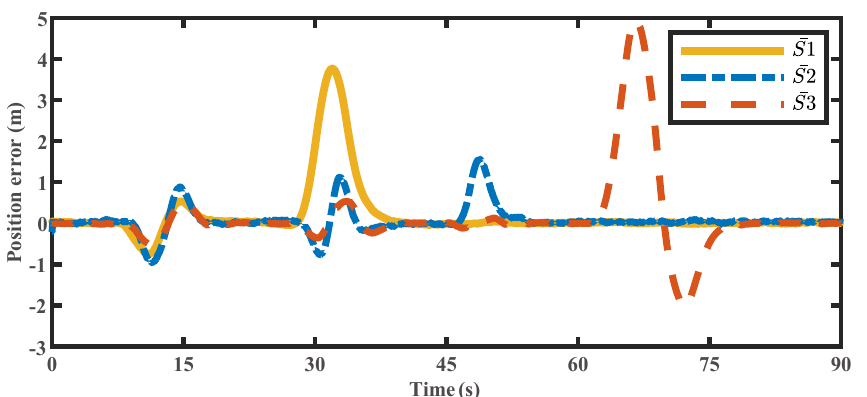}
    \caption{Position error of the platoon in HIL experiment.}
    \label{Fig20}
\end{figure}
Since the comprehensive control performance of the proposed MCF has been studied by numerical simulation in section III, the hardware-in-the-loop (HIL) experiment is conducted to further demonstrate the real time performance of the proposed MCF in this section. The layout and connection of the experimental platform is shown in Fig. 18. The MCF algorithm of one vehicle runs on the Speedgoat real-time target machine (Intel Core i7, 2.5 GHz), while the vehicle dynamic models and traffic scenario is implemented in the Concurrent real-time simulator (iHawk 45810, multi-core, RTX 3090 GPU) of Hirain test platform.

\par To verify the effectiveness of our method, the experiment is set the same as the obstacle avoidance scenario in section III without artificially-added communication delay, which will be explored in our future works. The platoon movement diagram of HIL experiment is shown in Fig. 19. As in the numerical simulation, each vehicle in the platoon can successfully avoid obstacle vehicles, and reunited as a platoon in the end. When there is blocking happened on the road, following vehicle 2 manages to balance the trade-off between driving safety and string-stability of the platoon with the proposed MCF.
\par Fig. 20 depicts the position error in the HIL experiment. It can be noted that the position error fluctuates during obstacle avoidance but eventually converges to uniform asymptotic stability. Since there is no artificially-add communication delay, the position error between each following vehicle can be controlled effectively and converges to 0. Furthermore, it should be noted that the average conducting time of the proposed method on Speedgoat real-time target machine is 1.1 milliseconds, which meets the real-time requirements. The HIL experiment demonstrates the effectiveness and real-time capability of the proposed MCF in critical scenarios.

\section{Conclusion}
In this paper, a holistic robust motion controller framework (MCF) was proposed to guarantee the safety of the platoon system. The longitudinal and lateral collaborative control problem of platooning was investigated. With the hierarchical structure, the H-infinity feedback controller was implemented in the UCL and the APF-based MPC was implemented in the LML. To summarize, comprehensive performance is considered in the holistic framework for autonomous platooning. It includes robustness, multi-objective control, obstacle avoidance ability, real-time capability and scalability. In order to validate the effectiveness of the MCF, three critical scenarios were constructed via the co-simulation of Simulink and TruckSim. The simulation results indicated that the platoon’s longitudinal string stability and obstacle avoidance capability could be efficiently realized by the MCF. And it can get a well improvement in multi-objective control performance compared with the single-structure MPC. Besides, the proposed MCF possessed a good scalability since that more robust controllers can be included in the hierarchical structure. Moreover, the HIL experiment was conducted to verify the effectiveness and real-time capability of the proposed MCF.
\par In the future work, the varieties of robust controllers and other SOTIF-related traffic regulations will be investigated and implemented in the holistic robust MCF. Furthermore, the corresponding real vehicle road tests will be carried out.

\ifCLASSOPTIONcaptionsoff
  \newpage
\fi



\bibliographystyle{IEEEtran}
\bibliography{ref}

\begin{thebibliography}{10}
\providecommand{\url}[1]{#1}
\csname url@samestyle\endcsname
\providecommand{\newblock}{\relax}
\providecommand{\bibinfo}[2]{#2}
\providecommand{\BIBentrySTDinterwordspacing}{\spaceskip=0pt\relax}
\providecommand{\BIBentryALTinterwordstretchfactor}{4}
\providecommand{\BIBentryALTinterwordspacing}{\spaceskip=\fontdimen2\font plus
\BIBentryALTinterwordstretchfactor\fontdimen3\font minus
  \fontdimen4\font\relax}
\providecommand{\BIBforeignlanguage}[2]{{%
\expandafter\ifx\csname l@#1\endcsname\relax
\typeout{** WARNING: IEEEtran.bst: No hyphenation pattern has been}%
\typeout{** loaded for the language `#1'. Using the pattern for}%
\typeout{** the default language instead.}%
\else
\language=\csname l@#1\endcsname
\fi
#2}}
\providecommand{\BIBdecl}{\relax}
\BIBdecl

\bibitem{lee2012development}
J.~Lee and B.~Park, ``Development and evaluation of a cooperative vehicle
  intersection control algorithm under the connected vehicles environment,''
  \emph{IEEE Transactions on Intelligent Transportation Systems}, vol.~13,
  no.~1, pp. 81--90, 2012.

\bibitem{zhao2016integrated}
Y.~Zhao, A.~Wagh, Y.~Hou, K.~Hulme, C.~Qiao, and A.~W. Sadek, ``Integrated
  traffic-driving-networking simulator for the design of connected vehicle
  applications: eco-signal case study,'' \emph{Journal of Intelligent
  Transportation Systems}, vol.~20, no.~1, pp. 75--87, 2016.

\bibitem{shladover1991automated}
S.~E. Shladover, C.~A. Desoer, J.~K. Hedrick, M.~Tomizuka, J.~Walrand, W.-B.
  Zhang, D.~H. McMahon, H.~Peng, S.~Sheikholeslam, and N.~McKeown, ``Automated
  vehicle control developments in the path program,'' \emph{IEEE Transactions
  on vehicular technology}, vol.~40, no.~1, pp. 114--130, 1991.

\bibitem{gao2016empirical}
S.~Gao, A.~Lim, and D.~Bevly, ``An empirical study of dsrc v2v performance in
  truck platooning scenarios,'' \emph{Digital Communications and Networks},
  vol.~2, no.~4, pp. 233--244, 2016.

\bibitem{hong2020joint}
C.~Hong, H.~Shan, M.~Song, W.~Zhuang, Z.~Xiang, Y.~Wu, and X.~Yu, ``A joint
  design of platoon communication and control based on lte-v2v,'' \emph{IEEE
  Transactions on Vehicular Technology}, vol.~69, no.~12, pp. 15\,893--15\,907,
  2020.

\bibitem{ju2020deception}
Z.~Ju, H.~Zhang, and Y.~Tan, ``Deception attack detection and estimation for a
  local vehicle in vehicle platooning based on a modified ufir estimator,''
  \emph{IEEE Internet of Things Journal}, vol.~7, no.~5, pp. 3693--3705, 2020.

\bibitem{wang2018review}
Z.~Wang, G.~Wu, and M.~J. Barth, ``A review on cooperative adaptive cruise
  control (cacc) systems: Architectures, controls, and applications,'' in
  \emph{2018 21st International Conference on Intelligent Transportation
  Systems (ITSC)}.\hskip 1em plus 0.5em minus 0.4em\relax IEEE, 2018, pp.
  2884--2891.

\bibitem{iso2019pas}
I.~ISO, ``Pas 21448-road vehicles-safety of the intended functionality,''
  \emph{International Organization for Standardization}, 2019.

\bibitem{konstantinopoulou2019specifications}
L.~Konstantinopoulou, A.~Coda, and F.~Schmidt, ``Specifications for multi-brand
  truck platooning,'' in \emph{ICWIM8, 8th International Conference on
  Weigh-In-Motion}, 2019, pp. 8--p.

\bibitem{wang2022robust}
B.~Wang, Y.~Luo, Z.~Zhong, and K.~Li, ``Robust non-fragile fault tolerant
  control for ensuring the safety of the intended functionality of cooperative
  adaptive cruise control,'' \emph{IEEE Transactions on Intelligent
  Transportation Systems}, 2022.

\bibitem{seiler2004disturbance}
P.~Seiler, A.~Pant, and K.~Hedrick, ``Disturbance propagation in vehicle
  strings,'' \emph{IEEE Transactions on automatic control}, vol.~49, no.~10,
  pp. 1835--1842, 2004.

\bibitem{li2019platoon}
Y.~Li, W.~Chen, S.~Peeta, and Y.~Wang, ``Platoon control of connected
  multi-vehicle systems under v2x communications: design and experiments,''
  \emph{IEEE Transactions on Intelligent Transportation Systems}, vol.~21,
  no.~5, pp. 1891--1902, 2019.

\bibitem{zhao2020vehicle}
C.~Zhao, X.~Duan, L.~Cai, and P.~Cheng, ``Vehicle platooning with non-ideal
  communication networks,'' \emph{IEEE transactions on vehicular technology},
  vol.~70, no.~1, pp. 18--32, 2020.

\bibitem{lan2021data}
J.~Lan, D.~Zhao, and D.~Tian, ``Data-driven robust predictive control for mixed
  vehicle platoons using noisy measurement,'' \emph{IEEE Transactions on
  Intelligent Transportation Systems}, 2021.

\bibitem{zhao2020stability}
C.~Zhao, L.~Cai, and P.~Cheng, ``Stability analysis of vehicle platooning with
  limited communication range and random packet losses,'' \emph{IEEE Internet
  of Things Journal}, vol.~8, no.~1, pp. 262--277, 2020.

\bibitem{mamduhi2020event}
M.~H. Mamduhi, E.~Hashemi, J.~S. Baras, and K.~H. Johansson, ``Event-triggered
  add-on safety for connected and automated vehicles using road-side network
  infrastructure,'' \emph{IFAC-PapersOnLine}, vol.~53, no.~2, pp.
  15\,154--15\,160, 2020.

\bibitem{xu2018stable}
L.~Xu, W.~Zhuang, G.~Yin, and C.~Bian, ``Stable longitudinal control of
  heterogeneous vehicular platoon with disturbances and information delays,''
  \emph{IEEE Access}, vol.~6, pp. 69\,794--69\,806, 2018.

\bibitem{zheng2016distributed}
Y.~Zheng, S.~E. Li, K.~Li, F.~Borrelli, and J.~K. Hedrick, ``Distributed model
  predictive control for heterogeneous vehicle platoons under unidirectional
  topologies,'' \emph{IEEE Transactions on Control Systems Technology},
  vol.~25, no.~3, pp. 899--910, 2016.

\bibitem{xu2017dsrc}
Z.~Xu, X.~Li, X.~Zhao, M.~H. Zhang, and Z.~Wang, ``Dsrc versus 4g-lte for
  connected vehicle applications: A study on field experiments of vehicular
  communication performance,'' \emph{Journal of advanced transportation}, vol.
  2017, 2017.

\bibitem{chen2020cooperative}
Y.~Chen, C.~Lu, and W.~Chu, ``A cooperative driving strategy based on velocity
  prediction for connected vehicles with robust path-following control,''
  \emph{IEEE Internet of Things Journal}, vol.~7, no.~5, pp. 3822--3832, 2020.

\bibitem{baldi2020establishing}
S.~Baldi, D.~Liu, V.~Jain, and W.~Yu, ``Establishing platoons of bidirectional
  cooperative vehicles with engine limits and uncertain dynamics,'' \emph{IEEE
  Transactions on Intelligent Transportation Systems}, vol.~22, no.~5, pp.
  2679--2691, 2020.

\bibitem{montemerlo2008junior}
M.~Montemerlo, J.~Becker, S.~Bhat, H.~Dahlkamp, D.~Dolgov, S.~Ettinger,
  D.~Haehnel, T.~Hilden, G.~Hoffmann, B.~Huhnke \emph{et~al.}, ``Junior: The
  stanford entry in the urban challenge,'' \emph{Journal of field Robotics},
  vol.~25, no.~9, pp. 569--597, 2008.

\bibitem{zheng2005evolutionary}
C.~Zheng, L.~Li, F.~Xu, F.~Sun, and M.~Ding, ``Evolutionary route planner for
  unmanned air vehicles,'' \emph{IEEE Transactions on robotics}, vol.~21,
  no.~4, pp. 609--620, 2005.

\bibitem{ji2016path}
J.~Ji, A.~Khajepour, W.~W. Melek, and Y.~Huang, ``Path planning and tracking
  for vehicle collision avoidance based on model predictive control with
  multiconstraints,'' \emph{IEEE Transactions on Vehicular Technology},
  vol.~66, no.~2, pp. 952--964, 2016.

\bibitem{liu2015predictive}
P.~Liu and {\"U}.~{\"O}zg{\"u}ner, ``Predictive control of a vehicle convoy
  considering lane change behavior of the preceding vehicle,'' in \emph{2015
  American Control Conference (ACC)}.\hskip 1em plus 0.5em minus 0.4em\relax
  IEEE, 2015, pp. 4374--4379.

\bibitem{yang2021comparative}
K.~Yang, X.~Tang, Y.~Qin, Y.~Huang, H.~Wang, and H.~Pu, ``Comparative study of
  trajectory tracking control for automated vehicles via model predictive
  control and robust h-infinity state feedback control,'' \emph{Chinese Journal
  of Mechanical Engineering}, vol.~34, no.~1, pp. 1--14, 2021.

\bibitem{huang2018review}
Y.~Huang, H.~Wang, A.~Khajepour, B.~Li, J.~Ji, K.~Zhao, and C.~Hu, ``A review
  of power management strategies and component sizing methods for hybrid
  vehicles,'' \emph{Renewable and Sustainable Energy Reviews}, vol.~96, pp.
  132--144, 2018.

\bibitem{rasekhipour2016potential}
Y.~Rasekhipour, A.~Khajepour, S.-K. Chen, and B.~Litkouhi, ``A potential
  field-based model predictive path-planning controller for autonomous road
  vehicles,'' \emph{IEEE Transactions on Intelligent Transportation Systems},
  vol.~18, no.~5, pp. 1255--1267, 2016.

\bibitem{gao2016robust}
F.~Gao, S.~E. Li, Y.~Zheng, and D.~Kum, ``Robust control of heterogeneous
  vehicular platoon with uncertain dynamics and communication delay,''
  \emph{IET Intelligent Transport Systems}, vol.~10, no.~7, pp. 503--513, 2016.

\bibitem{li2017distributed}
S.~E. Li, X.~Qin, Y.~Zheng, J.~Wang, K.~Li, and H.~Zhang, ``Distributed platoon
  control under topologies with complex eigenvalues: Stability analysis and
  controller synthesis,'' \emph{IEEE Transactions on Control Systems
  Technology}, vol.~27, no.~1, pp. 206--220, 2017.

\bibitem{pare2019networked}
P.~E. Par{\'e}, E.~Hashemi, R.~Stern, H.~Sandberg, and K.~H. Johansson,
  ``Networked model for cooperative adaptive cruise control,''
  \emph{IFAC-PapersOnLine}, vol.~52, no.~20, pp. 151--156, 2019.

\bibitem{moon2001delay}
Y.~S. Moon, P.~Park, W.~H. Kwon, and Y.~S. Lee, ``Delay-dependent robust
  stabilization of uncertain state-delayed systems,'' \emph{International
  Journal of control}, vol.~74, no.~14, pp. 1447--1455, 2001.

\bibitem{el1997cone}
L.~El~Ghaoui, F.~Oustry, and M.~AitRami, ``A cone complementarity linearization
  algorithm for static output-feedback and related problems,'' \emph{IEEE
  transactions on automatic control}, vol.~42, no.~8, pp. 1171--1176, 1997.

\bibitem{wu2006delay}
M.~Wu, Y.~He, and J.-H. She, ``Delay-dependent stabilization for systems with
  multiple unknown time-varying delays,'' \emph{International Journal of
  Control, Automation, and Systems}, vol.~4, no.~6, pp. 682--688, 2006.

\bibitem{wang2019crash}
H.~Wang, Y.~Huang, A.~Khajepour, Y.~Zhang, Y.~Rasekhipour, and D.~Cao, ``Crash
  mitigation in motion planning for autonomous vehicles,'' \emph{IEEE
  transactions on intelligent transportation systems}, vol.~20, no.~9, pp.
  3313--3323, 2019.

\bibitem{schulman2013finding}
J.~Schulman, J.~Ho, A.~X. Lee, I.~Awwal, H.~Bradlow, and P.~Abbeel, ``Finding
  locally optimal, collision-free trajectories with sequential convex
  optimization.'' in \emph{Robotics: science and systems}, vol.~9, no.~1.\hskip
  1em plus 0.5em minus 0.4em\relax Citeseer, 2013, pp. 1--10.

\bibitem{rasekhipour2018autonomous}
Y.~Rasekhipour, I.~Fadakar, and A.~Khajepour, ``Autonomous driving motion
  planning with obstacles prioritization using lexicographic optimization,''
  \emph{Control Engineering Practice}, vol.~77, pp. 235--246, 2018.

\bibitem{boggs1995sequential}
P.~T. Boggs and J.~W. Tolle, ``Sequential quadratic programming,'' \emph{Acta
  numerica}, vol.~4, pp. 1--51, 1995.

\bibitem{de2019stability}
F.~de~Oliveira~Souza, L.~A.~B. Torres, L.~A. Mozelli, and A.~A. Neto,
  ``Stability and formation error of homogeneous vehicular platoons with
  communication time delays,'' \emph{IEEE Transactions on Intelligent
  Transportation Systems}, vol.~21, no.~10, pp. 4338--4349, 2019.

\end{thebibliography}
%

%

\begin{IEEEbiography}[{\includegraphics[width=1in,height=1.25in,clip,keepaspectratio]{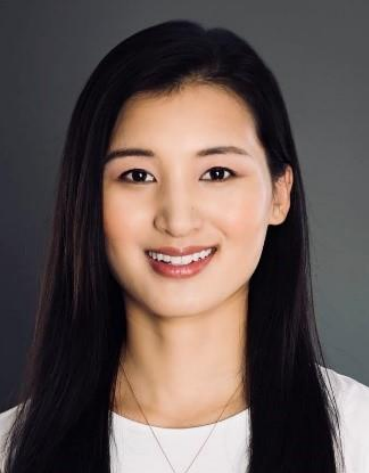}}]{Hong Wang}
is currently a Research Associate Professor at Tsinghua University. From the year 2015 to 2019, she was working as a Research Associate of Mechanical and Mechatronics Engineering with the University of Waterloo. She received her Ph.D. degree in Beijing Institute of Technology in China in the year 2015. Her research focuses on the safety of the on-board AI algorithm, the safe decision-making for intelligent vehicles, and the test and evaluation of SOTIF. She becomes the IEEE member since the year 2017. She has published over 60 papers on top international journals. She also served as the associate editor for 2019 Intelligent Vehicles Symposium held in Paris, France.
\end{IEEEbiography}

\begin{IEEEbiography}[{\includegraphics[width=1in,height=1.25in,clip,keepaspectratio]{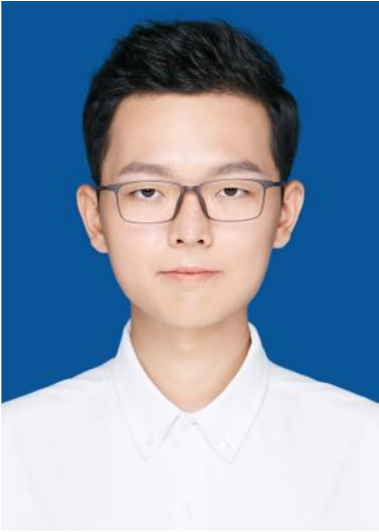}}]{Li-Ming Peng}
is currently pursuing a master’s degree in the department of Vehicle Engineering at Hefei University of Technology. He received the B.Eng. degree in Vehicle Engineering from Hefei University of Technology in 2019. His research focuses on decision-making and vehicle dynamics control for autonomous platooning, and the SOTIF of the intelligent and connected vehicle.
\end{IEEEbiography}

\begin{IEEEbiography}[{\includegraphics[width=1in,height=1.25in,clip,keepaspectratio]{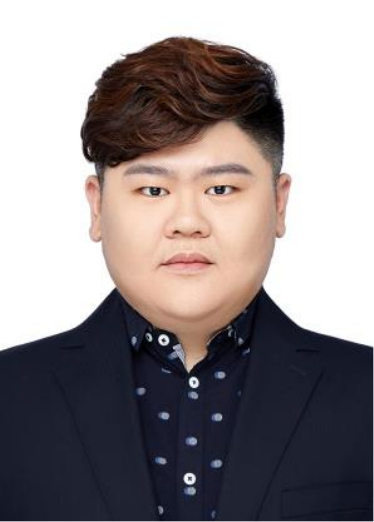}}]{Zichun Wei}
received the M.S. degree in Electrical Engineering from University of Southern California, Los Angeles, United States, 2020. He received the B.S. degree in Mechanical Engineering from Miami University, Oxford, United States, 2018.  His research interests include multi objective control, data driven control and motion planning for multi-agents systems.
\end{IEEEbiography}

\begin{IEEEbiography}[{\includegraphics[width=1in,height=1.25in,clip,keepaspectratio]{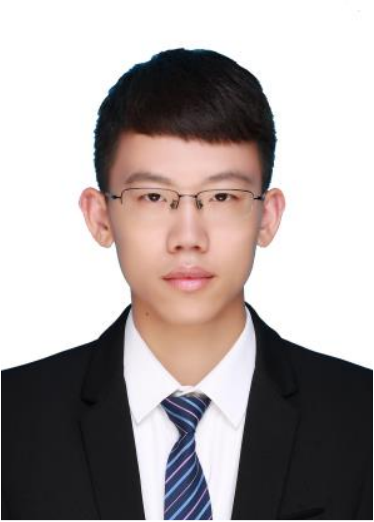}}]{Kai Yang}
is currently pursuing the Ph.D. degree in the Department of Automotive Engineering, Chongqing University, Chongqing, China. He received the B.E. degree in vehicle engineering from the Wuhan University of Technology in 2018. His research interests focus on decision-making
and vehicle dynamics control of autonomous vehicles.
\end{IEEEbiography}
\vspace{-2cm}
\begin{IEEEbiography}[{\includegraphics[width=1in,height=1.25in,clip,keepaspectratio]{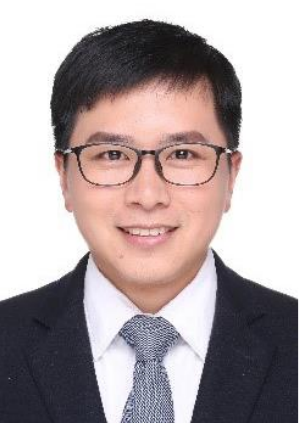}}]{Xian-Xu ‘Frank’ Bai}
is currently an Associate Professor with the Department of Vehicle Engineering, Hefei University of Technology, Hefei, China. He received his Ph.D. degree in Instrument Science and Technology from Chongqing University in 2013. He was a Joint Ph.D. student of Chongqing University and University of Maryland, College Park. His research focuses on new mechatronics-based vehicle dynamics and control with emphasis on intelligent/unmanned vehicles.
Currently, Dr. Bai serves as Associate Editor of the Journal of Intelligent Material Systems and Structures and SAE International Journal of Connected and Automated Vehicles. 
\end{IEEEbiography}
\vspace{-2cm}
\begin{IEEEbiography}[{\includegraphics[width=1in,height=1.25in,clip,keepaspectratio]{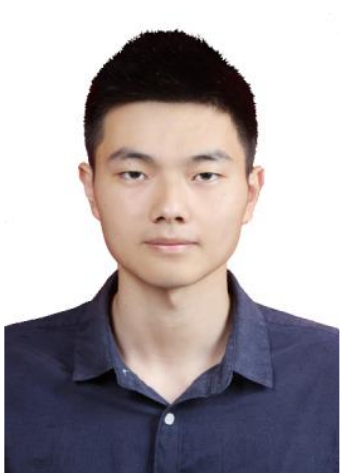}}]{Luo Jiang}
 is currently a Ph.D. student with the Department of Mechanical Engineering, University of Alberta, Edmonton, AB, Canada. He received the B.Sc. degree in Vehicle Engineering, and the M.Sc. degree in Mechatronic System Engineering from Southwest University, Chongqing, China, in 2017 and 2020, respectively. His research interests include safety and energy conservation of vehicle platooning.
\end{IEEEbiography}
\vspace{-2cm}
\begin{IEEEbiography}[{\includegraphics[width=1in,height=1.25in,clip,keepaspectratio]{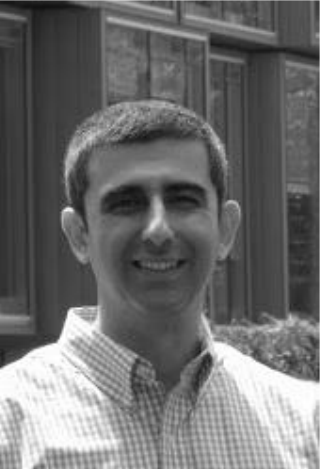}}]{Ehsan Hashemi}
received his PhD in Mechanical and Mechatronics Engineering in 2017 from University of Waterloo, ON, Canada, and is currently an Assistant Professor at the University of Alberta and director of the Networked Optimization, Diagnosis, and Estimation (NODE) lab. Previously, he was a visiting professor at the School of Electrical Engineering and Computer Science, KTH Royal Institute of Technology in 2019. His research is focused on distributed and fault-tolerant control, and cooperative intelligent transport system, and human-robot interaction.
\end{IEEEbiography}
\vspace{-1cm}






\end{document}